\documentclass[%
superscriptaddress,
nofootinbib,
 amsmath,amssymb,
 aps,
prb,
 twocolumn
]{revtex4-2}
\pdfoutput=1
\usepackage{graphicx}
\usepackage{dcolumn}
\usepackage{bm}
\usepackage{hyperref}

\usepackage{inputenc}
\usepackage{epsfig}
\usepackage{soul}

\newcommand{\bea}{\begin{eqnarray}}
\newcommand{\eea}{\end{eqnarray}}
\newcommand{\be}{\begin{equation}}
\newcommand{\ee}{\end{equation}}
\newcommand{\ba}{\begin{align}}
\newcommand{\ea}{\end{align}}

\def\Or[#1]{{\text{O}}\left({#1}\right)}
\def\dotl[#1,#2]{\left\langle #1, #2 \right\rangle}
\def\dotlb[#1,#2]{[ #1, #2 ]}
\def\dotp[#1,#2]{(#1) \cdot (#2)}
\def\aff[#1,#2]{\hat{#1}(#2)}
\def\n4sym{{\cal N}=4 SYM}
\def\>{\rangle}
\def\<{\langle}
\def\weight[#1,#2,#3]{\{(#1),#2,#3\}}
\def\ads[#1]{$\text{AdS}_{#1}$}

\usepackage{xcolor}

\newcommand{\ddt}{\frac{\mathrm d}{\mathrm dt}}
\newcommand{\dt}[1]{\frac{\mathrm d #1}{\mathrm dt}}

\newcommand{\vol}{{d^dx}}
\newcommand{\charge}{Q}
\newcommand{\chdens}{\rho}
\newcommand{\dipole}{Q}
\newcommand{\mom}{P}
\newcommand{\momdens}{p}
\newcommand{\coords}{t, \bm{x}}

\newcommand{\EMT}{T}
\newcommand{\hdens}{h}
\newcommand{\vel}{V}

\newcommand{\veci}{\vec{\textit{\i}}}
\newcommand{\vecj}{\vec{\textit{\j}}}

\newcommand{\eqendspace}{\,}

\newcommand{\bpsi}{\bar{\psi}}
\newcommand{\bpsisq}{\bar{\psi}^2}
\newcommand{\bchi}{\bar{\chi}}

\begin{document}

\title{Hydrodynamics of ideal fracton fluids}

\author{Kevin T. Grosvenor} 
\affiliation{Max Planck Institute for the Physics of Complex Systems, 01187 Dresden, Germany}
\affiliation{W\"{u}rzburg-Dresden Cluster of Excellence ct.qmat, Germany}
\author{Carlos Hoyos} 
\affiliation{Department  of  Physics  and  Instituto  de  Ciencias  y  Tecnolog\'{i}as  Espaciales  de  Asturias  (ICTEA)\\
Universidad  de  Oviedo,  c/  Federico  Garc\'{i}a  Lorca  18,  ES-33007  Oviedo,  Spain}
\author{Francisco Pe\~{n}a-Benitez} 
\affiliation{Max Planck Institute for the Physics of Complex Systems, 01187 Dresden, Germany}
\affiliation{W\"{u}rzburg-Dresden Cluster of Excellence ct.qmat, Germany}
\author{Piotr Sur\'owka} 
\affiliation{Department of Theoretical Physics, Wroc\l{}aw  University  of  Science  and  Technology,  50-370  Wroc\l{}aw,  Poland}
\affiliation{Max Planck Institute for the Physics of Complex Systems, 01187 Dresden, Germany}
\affiliation{W\"{u}rzburg-Dresden Cluster of Excellence ct.qmat, Germany}

\begin{abstract}
Low-energy dynamics of many-body fracton excitations necessary to describe topological defects should be governed by a novel type of hydrodynamic theory. We use a Poisson bracket approach to systematically derive hydrodynamic equations from conservation laws of scalar theories with fracton excitations.   We study three classes of theories. In the first class we introduce a general action for a scalar with a shift symmetry linear in the spatial coordinates, whereas the second one correspond with a complex scalar, while the third class serves as a toy model for disclinations and dislocations propagating along the Burgers vector.  We apply our construction to study hydrodynamic fluctuations around equilibrium states and derive the dispersion relations of hydrodynamic modes.
\end{abstract}

\maketitle


\section{Introduction}

Recent years have seen a dramatic growth of interest in the study of \emph{fracton} phases of matter \cite{nandkishore_fractons_2019,Pretko:2020cko}. In these phases, the motion of the elementary fractonic excitations is spatially restricted. While earlier spin models with such excitations (e.g. \cite{Chamon:2005,Haah2011,Bravyi:2011,Vijay:2016phm,Williamson2016}) describe \emph{gapped} fracton phases, \emph{gapless} phases have also been proposed in the context of certain spin liquids \cite{Xu2006,Xu2010,Pretko2017spinliquid,Pretko2017spinliquid2,You_emergent_2020}, dipole-conserving lattice models \cite{Pai2019,Feldmeier2020,Morningstar2020,Iaconis2021,moudgalya2021spectral} and quantum elasticity \cite{PretkoSolid2018,Gromov2019elastic,Kumar2019,Pretko:2018fed,pretko2019crystal,zhai2019two,Gromov:2019waa,Nguyen:2020yve,Fruchart2020dual,Manoj:2020abe,Surowka:2021ved} (see also \cite{Paramekanti2002,Sandvik2002,Rousseau2004}, where similar structures appear). From an effective field theory perspective, such gapless phases are of particular interest as they are typically the ones which survive and remain robust at very low energies. The dynamics of such a low-energy regime should be governed by a hydrodynamic theory, which reflects the symmetries and, in consequence, the conservation laws of a fractonic system.

Hydrodynamics offers a phenomenological description of interacting many-body systems. It focuses on a subset of physical quantities that remain conserved at low energies such as the particle number or momentum. Fractonic excitations conserve not only charge but also one or several higher moments of charge. It follows that the low-energy regime is characterized by long thermalization, sub-diffusive behavior of systems without momentum conservation, and a different structure of continuity equations. This has already been noticed in models of a scalar field with fracton excitations that emerge as a generalization of ordinary superfluids \cite{Yuan:2019geh,Chen2021} or in dissipative diffusive evolution of charges with restricted mobility \cite{GromovLucas2020}. Another class of fractonic fluids arises in the dynamics of topological defects \cite{Aasen:2020zru}.

Topological defects arise in many areas of physics and play an important role in, e.g., in superfluids \cite{sonin_dynamics_2016}, quantum Hall effect \cite{ezawa_quantum_2013}, liquid crystals \cite{Turner2010} and metamaterials \cite{Bertoldi2017}. The interpretation of topological defects as fractons is most visible in the language of elastic dualities \cite{kleinert1983double,Zaanen2004,Cvetkovic_2006,Beekman2017} that map elastic displacements to tensor gauge fields and the defects to charges that source the gauge fields. As was first observed in \cite{Pretko2017spinliquid,Pretko2017spinliquid2} fractonic charges with restricted mobility naturally source tensor gauge fields. Therefore topological defects can be interpreted as fractons. This interpretation allows one to construct low-energy coarse-grained models of defect dynamics based on the underlying symmetries, which are otherwise difficult to obtain and challenging to study from first principles, even numerically. Such an approach was successfully applied to superfluid vortices \cite{Wiegmann:2013hca,Watanabe2013,LucasSurowkaVortex,YuBradley2017,Moroz:2018noc,Bogatskiy2019,Wiegmann2019,Doshi2021}.

In this paper, our interest is in the effective field theory description of fracton fluids. For simplicity, we will only consider non-dissipative fluids and leave the study of additional phenomena, such as dissipation, for future work. We follow the approach of \cite{Landau1941,Dzyaloshinskii1980,Morrison1980,zakharov_hamiltonian_1997,Son:2019qlm}, starting from a sought-after algebra of symmetries, the corresponding conservation equations, and eventually deriving the hydrodynamic equations of motion for an ideal fluid. We focus on three classes of models: a real scalar theory with a shift symmetry linear in the spatial coordinates that extends previously studied models \cite{Gromov_multipole_2019,GromovLucas2020,Seiberg:2020bhn}, a complex scalar theory with a global $U(1)$ and a global vector symmetry previously studied in \cite{Pretko:2018jbi, Seiberg:2019vrp}, and a chiral scalar theory that has a Burgers-like vector, constraining the movement of charges. While the non-chiral models are analyzed in general dimensions, the proliferation of types of defects in spatial dimensions higher than two leads us to focus the analysis of the chiral model to two spatial dimensions. The chiral theory can be viewed as a toy model to understand the basic physical properties of many-body disclination dynamics. Given these theories we motivate the corresponding hydrodynamic Poisson structure and the derivative expansion in hydrodynamics. The main outcome of our analysis is that the Poisson structure for fracton systems is the same as in conventional fluids, but the constitutive relations change, thus leading to novel types of hydrodynamic theories. We analyze the corresponding hydrodynamic modes in both theories. Our construction paves the way for a systematic study of transport properties in fracton theories.


\section{Hydrodynamic equations from Poisson brackets}

Hydrodynamic equations represent the macroscopic conservation laws for quantities preserved during the time evolution of a physical system. Although conceptually simple, the precise form of these equations can be rather challenging to obtain. The difficulty may be associated with the non-linear terms or the proper inclusion of degrees of freedom in case of systems with multiple conserved quantities. Therefore, in order to circumvent these difficulties, it is convenient to use the Poisson bracket formulation of hydrodynamics (for a review, see \cite{zakharov_hamiltonian_1997}). This has been successfully applied to various systems and theories, e.g., to superfluids \cite{Landau1941,Dzyaloshinskii1980}, liquid crystals \cite{Stark2005} and more recently the fractional quantum Hall effect \cite{Son:2019qlm}. One starts by postulating the Poisson brackets between conserved densities. Such a postulate can be justified a posteriori by computing corresponding commutators in a microscopic model. As an instructive example we present a derivation of the conservation laws for a fluid with Galilean symmetries. The relevant degrees of freedom are the number density $\chdens (\bm{x})$, momentum density $\momdens_i (\bm{x})$, and entropy density $s(\bm{x})$. The Poisson brackets for these variables read
\begin{subequations}
\begin{align}
\{ \momdens _ i(\bm{x}) , \chdens (\bm{y})    \} &= - \chdens (\bm{x}) \partial _{x^i} \delta(\bm{x}-\bm{y}) \eqendspace,\\
\{ \momdens _ i(\bm{x}) , s (\bm{y})    \}&= -s (\bm{x}) \partial _{x^i} \delta(\bm{x}-\bm{y}) \eqendspace,\\
\{ \momdens _ i(\bm{x}) , \momdens_j (\bm{y})    \}&= -[ \momdens_j (\bm{x}) \partial _{x^i} 
+ \momdens_i (\bm{y}) \partial _{x^j} ] \delta(\bm{x}-\bm{y}) \eqendspace,
\end{align}\label{eq:Poisson}
\end{subequations}
with all other Poisson brackets vanishing. In the subsequent analysis we focus on the zero temperature fluid. The energy density in the fluid is a function of hydrodynamic variables $\hdens = \hdens ( \chdens, \bm{\momdens})$. This is a thermodynamic potential, whose differential is given by
\begin{equation}
d  \hdens =\mu d \chdens+ v^i d \momdens_i \eqendspace,
\end{equation}
where we have introduced the chemical potential $\mu$ and fluid velocity $\bm{v}$. Note that we consider a non-stationary fluid configuration. The equations of motion follow from the Poisson brackets of the hydrodynamic variables with the Hamiltonian\footnote{Note that many references use different conventions, which results in a time evolution given by $\{H,\cdot \}$.}
\begin{subequations} \label{eq_hydroeqsintro}
\begin{align}
\partial_t \chdens   &  = \{\chdens ,H\}=  - \partial_i (\chdens v^i) \eqendspace, \\
\partial _t \momdens_i &= \{ \momdens_i, H\}=  - \chdens \, \partial _i \mu - \partial _j (v^j \momdens_i)  - \momdens_j \partial _i v^j \eqendspace,
\end{align}
\end{subequations}
where $H =\int \vol \, \hdens (\chdens, \bm{\momdens})$. The momentum equation is not yet in the form of a conservation law. However, using the Legendre transform of pressure $dp= \chdens d \mu  + \momdens_i dv^i$, we can rewrite it as a momentum flux
\begin{align}
\partial _t \momdens_i =  - \partial_k (v^k \momdens_i) - \partial_i p &= - \partial _k (v^k \momdens_i + p \delta_{i}^{k} ) \notag \\
&\equiv  - \partial_k T^{k}{}_{i} \eqendspace.
\end{align}
As a result, momentum conservation defines the stress tensor. This is a general form of the hydrodynamic equation in a theory that conserves particle number and momenta. It is valid both for non-relativistic and relativistic fluids. In order to derive the final form of Euler's equations for a fluid that respects the Galilean symmetry, we need to impose appropriate constraints. At this point, we can directly supplement the Poisson brackets \eqref{eq:Poisson} with a specific Hamiltonian or with a generic one whose form is fixed by the appropriate symmetries and thermodynamics. For example in the Galilean fluid the momentum density is equal to the mass flux of particles
\begin{equation}\label{eq:constraint}
\momdens_i = m \chdens v_i \eqendspace,
\end{equation}
where $m$ is the mass of the fluid constituents. Equation \eqref{eq:constraint} can be rewritten as a constraint on the pressure thermodynamic potential
\begin{equation}
\frac{\partial p}{\partial v_i}= m v_i \frac{\partial p}{\partial \mu} \eqendspace.
\end{equation}
with a solution given by $p=p\left( \mu + \frac{1}{2}m v^2 \right)$.
We can interpret this new variable as the chemical potential in the frame moving with the fluid element $\tilde{\mu} =  \mu + \frac{1}{2}m v^2$. Using this result we can recover the usual form of Euler's equations. The stress tensor has now a manifestly symmetric form
\begin{equation}
T_{ij}= m \chdens v_i v_j + p \delta _{ij}
\end{equation}
and the energy density is equal to
\begin{equation}
\hdens = \mu \chdens + v_i \momdens_i -p = \frac{1}{2} m \chdens v^2+ \tilde{\mu} \chdens - p \eqendspace,
\end{equation}
where we can identify the kinetic energy and the internal energy.

As a final step we want to justify our postulate \eqref{eq:Poisson} in a microscopic theory. This was first done by Landau in the context of superfluid Helium \cite{Landau1941}. We start with a Galilean-invariant Lagrangian
\begin{equation}
\mathcal{L}= \bpsi \left( i \partial _t +\frac{\nabla ^2}{2m} \right )\psi \eqendspace,
\end{equation}
where $\psi$ is a complex scalar field. Since we need the commutator between density and momentum current we need to compute the Noether currents for $U(1)$ symmetry and translations. The symmetry transformation is given by
\begin{subequations} \label{eq:U(1)}
\begin{align}
\psi(t,\bm{x}) &\rightarrow e^{i \alpha} \psi (t,\bm{x}) \eqendspace, \\
\bpsi (t,\bm{x}) &\rightarrow e^{-i \alpha} \bpsi (t,\bm{x}) \eqendspace.
\end{align}
\end{subequations}
We promote the parameter $\alpha$ to be a function of space and time and perform a variation of the action with respect to it
\begin{align}\label{eq:relax}
 \delta S &= \alpha (t,\bm{x})\bigg[\partial_t( \bpsi (t,\bm{x}) \psi(t,\bm{x}))\\  \nonumber
 &+ \frac{i}{2m} \Bigl( \nabla ^2 \bpsi (t,\bm{x}) \psi(t,\bm{x})- \bpsi (t,\bm{x})  \nabla ^2\psi(t,\bm{x}) \Bigr) \bigg] \eqendspace.  \nonumber
\end{align}
We can now identify the density
\begin{equation}
\chdens (t,\bm{x}) = \bpsi (t,\bm{x}) \psi(t,\bm{x})
\end{equation}
and the particle current density from which we can compute the momentum density
\begin{equation}
\momdens_i= \frac{i}{2} \left[\nabla _i \bpsi (t,\bm{x}) \psi(t,\bm{x})- \bpsi (t,\bm{x})  \nabla _i \psi(t,\bm{x})\right] \eqendspace.
\end{equation}
Taking $\{\psi (\bm{x}), \bpsi (\bm{y})\}=\delta (\bm{x}-\bm{y})$, we can derive Eqs. \eqref{eq:Poisson}. In our analysis of the fracton fluids, we will employ the same procedure by postulating the brackets and justifying them using microscopic models.


\section{Monopole-Dipole-Momentum Algebra}

Our analysis will be phrased in the language of multipole algebras constructed in \cite{Gromov_multipole_2019}. The basic ingredient of this construction is the simultaneous conservation of the fracton monopole charge $\charge$ and the dipole moment $\dipole^i$, $i= 1, \ldots , d$ ranging over the spatial directions. These are defined to be the spatial integrals of their respective densities:
\begin{subequations}
\begin{align}
    & \dt{\charge} = \ddt \int \vol \, \chdens = 0\eqendspace, \\
    & \dt{\dipole^i} = \ddt \int \vol \, x^i \chdens = 0\eqendspace.
\end{align}
\end{subequations}
where $\chdens ( \coords )$ is the fracton charge density.

A sufficient condition to guarantee these conservation laws is a generalized continuity equation of the form
\begin{equation}\label{eq:continuity}
    \partial_t \chdens + \partial_i \partial_j J^{ij} = 0\eqendspace.
\end{equation}
In addition, we assume that the system is translationally invariant, which implies the conservation of momentum
\begin{equation}
    \dt{\mom_i} = \ddt \int \vol \, \momdens_i = 0\eqendspace.
\end{equation}
The charges $\charge$, $\dipole^i$, and $\mom_i$ naturally act on the system as generators of the symmetry group. Fields $\Phi$ in a given representation will transform according to $\delta_{\alpha}\Phi = \{\Phi,\alpha \charge\}$, $\delta_{\beta}\Phi = \{\Phi,\beta_i \dipole^i\}$, and $\delta_{\gamma}\Phi =\{\Phi, \gamma^i \mom_i\}$, where $\alpha$, $\beta_i$, and $\gamma^i$ are constant infinitesimal parameters. For example, these symmetries can be represented by a real scalar field $\phi ( \coords )$ transforming as
\begin{subequations} \label{eq:phitrans}
\begin{align}
    & \delta_{\alpha} \phi ( \coords ) = \alpha\eqendspace, \\
    & \delta_{\beta} \phi ( \coords ) = \beta_i x^i\eqendspace, \\
    & \delta_{\gamma} \phi ( \coords ) = \gamma^i \partial_i \phi ( \coords )\eqendspace. \label{eq:momtrans1}
\end{align}
\end{subequations}
Another example is that of a complex scalar field $\psi ( \coords )$ transforming under a global vector transformation
\begin{equation} \label{eq:globalvector}
    \psi ( \coords ) \rightarrow e^{i \boldsymbol{\beta} \cdot \boldsymbol{x}} \psi ( \coords ).
\end{equation}
in addition to the usual $U(1)$ transformation in Eqn. \eqref{eq:U(1)} and the same momentum transformation as for the real scalar in Eqn. \eqref{eq:momtrans1}. In fact, the real scalar $\phi$ transforms precisely like the phase of the complex scalar field $\psi$. Indeed, the low-energy limit of a theory of $\psi$ in the spontaneously broken phase, in which the global $U(1)$ and vector symmetries are broken, is described by the corresponding Nambu-Goldstone boson, which is precisely the phase $\phi$ of $\psi$. However, it must be noted that not all theories of $\phi$ with the symmetries \eqref{eq:phitrans} necessarily arise in this way.

The monopole transformation, $\charge$, is nothing but the constant shift symmetry familiar in the context of Nambu-Goldstone bosons. The dipole transformation, $\dipole^i$, is just the spatial part of the linear shift symmetry familiar in the context of Galileons \cite{Nicolis:2008in}. In general, one can consider the so-called polynomial shift symmetry whose charges are generated by polynomials in the spatial coordinates of any finite degree \cite{Griffin:2014bta}. An example of such a scenario that has been studied previously is the traceless scalar theory \cite{Pretko:2016lgv}, which in addition to the conserved charges $\charge,\charge^i$ also conserves the second moment $\charge^{(2)}=\int\vol ||\mathbf x||^2\chdens$.

The form of the Monopole-Dipole-Momentum algebra (MDMA) is as follows,
\begin{subequations}
\begin{align} \label{eq:commqipi}
	& \{ \charge , \charge \} = \{ \charge , \mom_i \} = \{ \charge , \dipole^i \} = 0\eqendspace, \\
	& \{ \mom_i, \dipole^j \} = \delta^j_i \charge \eqendspace.
\end{align}
\end{subequations}
Notice that momentum and dipole charge are charged with respect to each other. In fact,  if we shift the origin of the coordinates system in some direction ($\boldsymbol \gamma$),  its dipole moment will change accordingly to $\delta_{\gamma} Q^i = \gamma^i Q$.  On the other hand,  it is less intuitive that a dipole transformation with parameter $\boldsymbol \beta$ likewise modifies the momentum.  Nevertheless, the algebra \eqref{eq:commqipi} dictates that it does it according to $\delta_\beta \mom_i =\beta_i\charge$, which can be written in terms of the densities as follows
\begin{align}\label{eq:momtrans}
\delta_\beta\momdens_i &=\beta_i\chdens \,.
\end{align}

In the next sections we will focus on the hydrodynamic description of systems that are invariant under the whole symmetry group. Then, we apply the Poisson bracket formalism, to write down hydrodynamic equations for fractons. 
In particular, we will study two different classes of fracton theories that we dub chiral and non-chiral. However, the structure of hydrodynamic equations appears to be universal. The Poisson brackets \eqref{eq:Poisson} do not change, but the final form of the constitutive relations strongly depends on the Hamiltonian. The main distinguishing feature between the non-chiral and chiral theories is that the former will be assumed to be time-reversal invariant whereas in the later case we will allow for the breaking of time-reversal invariance while preserving a combination of time-inversions and a $\pi$ rotation.

In our analysis we do not introduce an additional independent charge for the dipole symmetry because the dipole charge is a derived quantity from the ordinary scalar charge. The situation is analogous to the orbital angular momentum or, in conformal theories, the charges associated to scale and conformal transformations. Those are not assigned independent charges, but are constructed from the components of the energy-momentum tensor. Their conservation follows from the properties of the energy-momentum tensor, conservation, symmetry and tracelessness. Similarly, the conservation of the dipole charge follows from the properties of the charge current, which is a total derivative of a symmetric tensor. One might consider the presence of an {\em intrinsic} dipole charge with a separate charge density, but we would not pursue this possibility here.


\section{Hydrodynamics of non-chiral fractons}

We start with theories preserving time-reversal invariance. We will do first a general analysis of hydrodynamics based on symmetries alone. In the hydrodynamic description we have to postulate the form of the Poisson brackets between conserved charges, based on the action they are expected to have as symmetry generators. On the other hand, the dependence of the Hamiltonian on the charges is constrained by the symmetries. Combining these two together, we are able to derive the constitutive relations for the currents in terms of the conserved charge and momenta. We do not make any assumption about the underlying microscopic theory, other than the symmetries. Once we derive the hydrodynamic equations and constitutive relations, we proceed to confirm our assumptions about the Poisson brackets and the hydrodynamic equations that follow from them by computing them in simple models where the symmetries are realized explicitly. 

For the present case, we find it convenient to introduce the generalized velocity $\vel_i=\chdens^{-1}\momdens_i$ which transforms as $\delta_\beta \vel_i=\beta_i$ (see Eq. \eqref{eq:momtrans}), and constrain the Hamiltonian to depend on the momentum only via the $\beta$-invariant combination $\partial_i \vel_j$. In fact, notice that a linear shift acts similarly to a Galilean boost, since the generalized velocity is shifted by the `boost' parameter $ \beta_i$. Therefore, we postulate that the low-energy effective Hamiltonian of the system have the following form
\begin{equation}
    H = \int \vol\, \hdens ( \chdens , \partial_i \vel_j , \partial_i \chdens ) +\mathcal O(\partial^3)\eqendspace.
\end{equation}
Since the independent dynamical variables of the system are $\chdens$ and $\momdens_i$, the variation of the Hamiltonian will have the form\footnote{Notice that we are not considering the dipole density $\rho^i(t,\mathbf{x})=x^i\rho(t,\mathbf{x})$ as a hydrodynamical variable since its conservation follows from the conservation of the charge density $\rho(t,\mathbf{x})$.}
\begin{equation}
    \delta H = \int\vol\,\left(\mu\delta\chdens+v^i\delta \momdens_i \right)\eqendspace,
\end{equation}
where the out-of-equilibrium chemical potentials take the form
\begin{subequations}
\begin{align}
    \label{eq_momentumchempot}  v^i  &= - \frac{1}{\chdens} \partial_j \frac{\partial \hdens}{\partial ( \partial_j \vel_i )} \eqendspace,\\
    \label{eq_chargechempot}  \mu &= \frac{\partial \hdens}{\partial \chdens} - \partial_i \frac{\partial \hdens}{\partial ( \partial_i \chdens )} - \frac{\momdens_i v^i}{\chdens}\eqendspace.
\end{align}
\end{subequations}
In addition, we postulate the the same Poisson brackets \eqref{eq:Poisson} as for ordinary fluids. In fact, in the next section we shall verify for a certain class of theories with linear shift symmetry that our assumptions are correct. We now derive the equations of motion for the conserved densities using $\partial _t S= \{S, H\}$, which are the same as Eqs. \eqref{eq_hydroeqsintro}. However, in the present case $\chdens v^i$ is a gradient as expected for a fractonic system. Therefore, the constitutive relation for the fractonic current read
\begin{equation}
	J^{ij} = -   \frac{\partial \hdens}{\partial ( \partial_{(j} \vel_{i)} )}\,.
\end{equation}
On  the other hand we can read out the momentum conservation equation the stress tensor
\begin{equation}
	T^i\,_j = p\delta^i_j  + v^i p_j  +  \frac{\partial \hdens}{\partial ( \partial_i \chdens )} \partial_j \chdens + \frac{\partial \hdens}{\partial ( \partial_i \vel_k )} \partial_j \vel_k \,,
\end{equation}
with the pressure $p$ defined as $p=\rho\mu + p_iv^i -h$. 

To be more precise, and understand the consequences of this proposal, we consider a time-reversal invariant quadratic Hamiltonian which would capture the dynamics of linear perturbations around the equilibrium state $\chdens = \chdens_0$, $\momdens_i =0$. The most general such Hamiltonian density reads
\begin{align}\label{eq:quadhydroHam}
\hdens =& \frac{\mu_0}{2 \chdens_0} \chdens^2  +\frac{1}{2 \chdens_0}\mu_1^{ij} \partial_i \chdens \, \partial_j \chdens  +\frac{\chdens_0}{2} v_1^{ijkl} \partial_i \vel_j \, \partial_k \vel_l \eqendspace,
\end{align}
with the phenomenological susceptibilities being positive definite in order that the Hamiltonian be bounded from below. Then, we fix the background density $\chdens=\chdens_0$ and pick the `frame' $\momdens_i=0$. In particular, the out-of-equilibrium chemical potentials $\mu$ and $v^i$ are
\begin{align}
\mu &= \chdens_0^{-1} ( \mu_0 \chdens - \mu_1^{ij} \partial_i \partial_j \chdens )\eqendspace,\\
v^j & = -v_1^{ijkl} \partial_i \partial_k \vel_l \eqendspace,
\end{align}
which imply the following set of linearized hydrodynamic equations
\begin{align}
\partial_t\delta\chdens - v_1^{ijkl}\partial_i\partial_j\partial_k \delta \momdens_l &= 0\eqendspace,\\
\partial_t\delta  \momdens_i + \left(\mu_0 -\mu_1^{jk}\partial_j\partial_k\right)\partial_i\delta\chdens&=0\eqendspace.
\end{align}
After Fourier transforming the system of equations, it is possible to verify the existence of two modes: one of them is non-propagating (at $\omega = 0$)\footnote{In general, we expect that dissipative terms will modify the dispersion relation of the non-propagating mode.}, while the propagating mode has the following dispersion relation
\begin{equation}\label{eq:dispersionsimple}
\omega^2 = \mu(q)v_1^{ijkl}q_iq_jq_kq_l \eqendspace,
\end{equation}
where $\mu(q)=\mu_0+\mu_1^{ij}q_iq_j$. Note that the positivity conditions required of all the susceptibilities appearing in Eq. \eqref{eq:quadhydroHam} guarantee that $\mu(q)v_1^{ijkl}$ is positive definite and, therefore, that the mode with dispersion \eqref{eq:dispersionsimple} will be a stable propagating mode. Also notice that the antisymmetric components $v_1^{[ij][kl]}$ are only relevant at the non-linear level, since they do not appear in the linearized equations of motion.


\subsection{Real Scalar field with linear shift symmetry}

In this section we will confirm the Poisson algebra and structure assumed in the hydrodynamic Hamiltonian for the fracton system with the linear shift symmetry \eqref{eq:phitrans}. In particular we shall assume theories with Lagrangian of the form
\begin{equation} \label{eq:L}
	L=\int\vol\,\mathcal L(\partial_t\phi,\partial_i\partial_j\phi)\eqendspace.
\end{equation}
For simplicity and clarity of presentation, we will consider this restricted functional dependence of the Lagrangian on the derivatives of the scalar field. Although \emph{a priori} this theory does allow for chiral modes we impose an additional symmetry requirement that they are absent. The generalization to Lagrangians depending on higher-order spatial derivatives of the scalar field is quite straightforward, but leads only to a proliferation of terms with no significant added insight. Such a form of the Lagrangian ensures that $\mathcal{L}$ by itself is invariant under the linear shift symmetry and not just its spatial integral $L$. We can allow $\mathcal{L}$ to vary by a total derivative, in which case it can depend on the first spatial derivative $\partial_i \phi$ as long as we insist that $\frac{\partial \mathcal{L}}{\partial ( \partial_i \phi )}$ is itself the gradient of some expression. Again, it is easy to handle adding this extra functional dependence in Eq. \eqref{eq:L}, but it does not lead to any added insight, at least insofar as the present analysis is concerned. Thus, we choose to restrict to the form \eqref{eq:L} (See appendix for the more general case).

After taking variations of the  Lagrangian we obtain
\begin{align}
	\delta L &=-\int \vol\left(\partial_t\pi+\partial_i\partial_jJ^{ij}\right)\delta\phi \notag \\
	&\quad + \int \vol\left(\partial_t(\pi\delta\phi)+\partial_i(\partial_jJ^{ij}\delta\phi-J^{ij}\partial_j\delta\phi\right)\eqendspace,
\end{align}
where we have defined
\begin{align} \label{eq:piandJ}
    \pi &= \frac{\partial \mathcal L}{\partial(\partial_t\phi)}\eqendspace, &%
    J^{ij} &= - \frac{\partial \mathcal L}{\partial(\partial_i\partial_j\phi)}\eqendspace.
\end{align}
For simplicity, we assume that the time derivative of the field can be obtained from $\pi$ by inverting with some functional $F$
\begin{equation}
    \partial_t\phi=F(\pi,\partial_i\partial_j\phi)\eqendspace.
\end{equation}
In the next section we will consider the example of a theory where this is not possible.
From the variation of the action we can simultaneously read the equations of motion and the Noether currents associated with the constant and linear shift. In fact, the equation of motion reads
\begin{align} \label{eq:eom}
    \partial_t \pi + \partial_i \partial_j J^{ij} &=0\eqendspace,
\end{align}
which can be identified with the monopole charge conservation with $\chdens=\pi$. In addition, the dipole density defined as $\chdens^i=x^i\pi$ is also conserved
\begin{align} \label{eq:ctyeqs}
	\partial_t \chdens^k + \partial_i j^{ik} &= 0 \eqendspace,
\end{align}
where
\begin{align} \label{eq:shiftjrels}
    j^{ik} &= x^k \partial_j J^{ij} - J^{ik}\eqendspace.
\end{align}
The system is also translationally invariant in time and space, which implies the conservation of energy and momentum. The energy density and flux are 
\begin{subequations}
\begin{align}
    \hdens &= \pi \, \partial_t \phi - \mathcal{L} (F, \partial_i \partial_j \phi ) \eqendspace,  \label{eq:Ttt} \\
    \EMT^{i}{}_{t} &= ( \partial_j J^{ij} ) \partial_t \phi - J^{ij} \partial_j \partial_t \phi \eqendspace, \label{eq:Tit}
\end{align}
\end{subequations}
while the momentum density and stress tensor are
\begin{subequations}
\begin{align}
    \momdens_i &=T^t\,_i= - \pi \partial_i \phi \eqendspace,\\
    \EMT^{j}{}_{i}  &=- \partial_k J^{jk} \partial_i\phi + J^{jk} \partial_k \partial_i \phi + \delta^j_i \mathcal L(F,\partial_i \partial_j \phi)\eqendspace, \label{eq:Tji}
\end{align}
\end{subequations}
They satisfy the usual conservation equations
\begin{subequations} \label{eq:EMcons}
\begin{align}
    \partial_t \hdens + \partial_i \EMT^{i}{}_{t} &= 0 \eqendspace, \\
    \partial_t \momdens_i + \partial_j \EMT^{j}{}_{i} &= 0 \eqendspace.
\end{align}
\end{subequations}

The definition of the momentum and charge densities for this class of systems automatically satisfy the transformation rule \eqref{eq:momtrans} and the Hamiltonian density \eqref{eq:Ttt} can be written as a functional of $\chdens$ and $\partial_i \vel_j$
\begin{equation}
h \equiv h( \chdens,\partial_i V_j) \eqendspace,
\end{equation}
as expected.
Finally, for this class of theories it is not hard to verify that the canonical bracket
\begin{equation} \label{eq:canonicalbracket}
    \{\phi(\bm{x}) , \pi(\bm{y})\} = \delta (\bm{x}-\bm{y})\eqendspace.
\end{equation}
implies the usual algebra \eqref{eq:Poisson} used above to derive the hydrodynamic theory.


\subsection{Complex Scalar field with linear shift symmetry}

Imposing the global vector symmetry \eqref{eq:globalvector} on a theory of a complex scalar field $\psi$ forbids an ordinary kinetic term that is quadratic in $\psi$ and contains some number of spatial derivatives. Instead, the theory of this kind that has been considered previously in \cite{Pretko:2018jbi, Seiberg:2019vrp} reads
\begin{align}
    \mathcal{L} &= | \partial_t \psi |^2 - m^2 | \psi |^2 - \frac{\lambda}{4} \bigl( | \psi |^2 \bigr)^2 - c_1 \partial_i | \psi |^2 \, \partial_i | \psi |^2 \notag \\
    &\quad - c_2 | \chi_{ij} |^2 - c_3 \bigl[ ( \psi^* )^2 \chi + \text{h.c.} \bigr],
\end{align}
where h.c. is the Hermitian conjugate,
\begin{equation}
    \chi_{ij} \equiv \psi \, \partial_i \partial_j \psi - \partial_i \psi \, \partial_j \psi,
\end{equation}
and $\chi = \chi_{ii}$ is its trace.  The canonical momenta of this system are $\pi_\psi = \partial_t \bpsi$, and $\pi_{\bpsi} = \partial_t \psi$.  For such theory the $U(1)$ Noether's charge $\rho$ and current  $J^i \equiv \partial_j J^{ij}$ densities are
%
\begin{subequations}
\begin{align}
    \rho &= i \bigl( \bpsi \partial_t \psi  -  \text{h.c.}\bigr), \\
    J_{ij} &= ic_2 \bigl( \bpsisq \chi_{ij} - \text{h.c.} \bigr),
\end{align}
\end{subequations}
%

On the other hand, the translational invariance of the system implies conservation of momentum. In particular,  the  momentum and stress densities are
\begin{subequations}
\begin{align}
    p_i &=- \left( \partial_i \bpsi \,  \pi_\psi + \text{h.c.} \right), \\
    T^{j}{}_{i} &=  \mathcal{L} \delta_{i}^{j} + 2c_1 \partial_j |\psi|^2 \partial_i |\psi|^2   \\
    &\quad + c_2 \Bigl( \bchi_{jk} \bigl( \chi_{ki} - 2 \partial_k \psi \, \partial_{i} \psi  \bigr) -  \frac{1}{2}\partial_k \bchi_{jk}  \partial_{i} \psi^2+  \text{h.c.} \Bigr) \notag \\
    &\quad + c_3 \Bigl( \bpsisq \bigl( \chi_{ij} - 2 \partial_j \psi \, \partial_{i} \psi \bigr)  - 2|\psi|^2 \partial_j \bpsi \, \partial_{i} \psi +  \text{h.c.} \Bigr)\notag.
\end{align}
\end{subequations}
In addition,  the system's Hamiltonian and energy current read
\begin{subequations}
\begin{align}
   h &= 2 |\partial_t\psi |^2 - \mathcal{L},\\
    T^{i}{}_{t} &= -2c_1 \partial_t |\psi|^2 \partial_i |\psi|^2   \\
    &\quad - c_2 \Bigl( \bchi_{ij} \bigl( \psi \, \partial_j \pi_{\bpsi} - 3 \partial_j \psi \, \pi_{\bpsi} \bigr) - \partial_j \bchi_{ij} \psi \pi_{\bpsi} + \text{h.c.}\Bigr)   \notag \\
    &\quad - c_3 \Bigl( \bpsisq \bigl( \psi \, \partial_i \pi_{\bpsi}- 3 \partial_i \psi \, \pi_{\bpsi} \bigr)- 2|\psi|^2  \partial_i \bpsi \, \pi_{\bpsi} + \text{h.c.} \Bigr)  \notag.
\end{align}
\end{subequations}

Using the equations of motion for this theory,  one can verify that conservation Eqs. \eqref{eq:continuity} and \eqref{eq:EMcons} are satisfied. Furthermore, the same canonical brackets as in Eqn. \eqref{eq:canonicalbracket} for $\psi$, $\bpsi$ and their conjugate momenta, imply the usual algebra between the conserved  densities Eqs. \eqref{eq:Poisson}.

In the broken phase, in which $| \psi |$ picks up a vacuum expectation value $\langle | \psi | \rangle$, at energies well below the mass scale, $E \ll m$, the theory just contains the Nambu-Goldstone mode, which is the phase $\phi$ of $\psi$. After rescaling time and space by a factor of $\langle | \psi | \rangle$, the Lagrangian is simply given by a free massless $z=2$ real scalar field theory: $\mathcal{L} = \frac{1}{2} ( \partial_t \phi )^2 - \frac{1}{4} c_2 ( \partial_i \partial_j \phi )^2$, which is a simple special case of the Lagrangian \eqref{eq:L} in the previous example.


\section{Hydrodynamics of chiral fractons}

In cases of physical interest, fractons may not only be restricted to move in a subdimensional space, but the direction of motion may be determined as well. An example of this are edge dislocations under an applied stress, whose direction of motion is determined by the Burgers vector ${\bm b}$ and the orientation of the defect.  Dual elastic fields couple to the disclination density $\chdens$ and current tensor $J^{ij}$ with a conservation equation  \eqref{eq:continuity} \cite{Pretko:2018fed}. In this system, the monopole charge $Q$ and dipole moment $Q^i$ can be identified with the disclination and dislocation charge respectively. In a two-dimensional lattice, the contribution  to the current tensor of a dislocation moving with velocity ${\bm v}$ takes the form \cite{Pretko:2018fed}
\be \label{eq:JBurger1}
J^{ij}\propto \epsilon^{(ik} b_k v^{j)} \eqendspace.
\ee
In the absence of vacancies, the current tensor is traceless $J^{i}{}_{i}=0$ \cite{Cvetkovic_2006,PhysRevLett.41.121}, forcing the velocity to be parallel to the Burgers vector $v^i= b^i v_b$. 

Let us introduce the unit vector $\hat{\bm b}={\bm b}/||{\bm b}||$ and define the coordinates parallel and transverse to the Burgers vector $b=\hat{b}_i x^i$, $s=\epsilon^{ij}x_i \hat{b}_j$, as well as the corresponding derivatives
\be
\partial_b \equiv \hat{b}^i \partial_i \eqendspace,\ \ \partial_s\equiv \epsilon^{ij} \hat{b}_j \partial_i \eqendspace.
\ee
With these conventions and assuming no vacancies the current is
\be \label{eq:JBurger2}
J^{bb}=J^{ss}=0 \eqendspace,\  \ J^{bs}=J^{sb}\propto \frac{1}{2}v_b \eqendspace. 
\ee

We expect that this to be a general form for the constitutive relation of the current. However, there are some subtleties regarding the realization of symmetries, as the breaking of rotational invariance allows also for subsystem symmetries. In this case we will follow a different path than for the non-chiral models. We will start by studying a simple model with the right symmetry realization and use it to derive the constitutive relations for the currents and hydrodynamic equations. Once expressed in terms of conserved charges, one could in principle forget about the microscopic origin and postulate the result as a hydrodynamic description for this class of models. However, it is possible that the constitutive relations derived in this form are not the most general, although we expect the general structure to be correct, some of the values of the coefficients might be allowed to change.

This structure can be captured by a simple model with a scalar field. The action reads
\be
S = \int d^3 x \, \mathcal L = \frac{1}{2}\int d^3 x \, \partial_s\phi \left( \partial_t \phi+\partial_b \phi\right) \eqendspace.
\ee
The equation of motion is
\be
\partial_t\partial_s \phi+\partial_b\partial_s \phi=0 \eqendspace.
\ee
This takes the form of the continuity equation \eqref{eq:continuity} if we identify
\be
\chdens \propto \partial_s\phi \eqendspace,\ \ J^{ss}=J^{bb}=0 \eqendspace,\ \ J^{bs} \propto \frac{1}{2}\phi \eqendspace,
\ee
where the proportionality constants are the same for both $\rho$ and $J^{bs}$. The energy, momentum and charge densities  in the Lagrangian formalism are
\begin{subequations}\label{eq:chdensities}
\begin{align}
    &e=T^t_{\ t}=\frac{\delta S}{\delta\partial_t \phi}\partial_t\phi- \mathcal L=-\frac{1}{2}\partial_s\phi \, \partial_b\phi \eqendspace, \\
    &k_i=T^t_{\ i}=-\frac{\delta S}{\delta\partial_t \phi}\partial_i\phi=-\frac{1}{2}\partial_s \phi \, \partial_i \phi \eqendspace, \\
    &n=\frac{\delta S}{\delta \partial_t \phi}=\frac{1}{2}\partial_s \phi \eqendspace.
\end{align}
\end{subequations}
Therefore, the model captures the structure of Eq. \eqref{eq:JBurger2} for dislocations of a fixed orientation. Dislocations of different orientations would be described by discrete spatial rotations of the action above. 

This model actually has a much larger symmetry than the one implied by the monopole, dipole and second moment conservation. The action transforms by a total derivative under the transformations
\be
\phi\to \phi+f(s)+g(t,b) \eqendspace,
\ee
which are also symmetries of the equation of motion. This implies that there is an infinite set of conserved charges
\be
Q(f,g)=\int d^2x\,(f(s)+g(t,b)) \, n \eqendspace. 
\ee
This model will also have blueuced conformal symmetry as those studied in \cite{Karch:2020yuy}. The $s$-independent transformation leaves $q$ and $k_s$ invariant but changes the momentum $k_b$ as follows
\be\label{eq:symmtranspb}
\delta k_b= - n \, \partial_b g(t,b).
\ee

Moving on to the Hamiltonian formalism, the canonical momentum conjugate to the scalar field is
\be
\pi=\frac{\delta S}{\delta \partial_t \phi}=\frac{1}{2}\partial_s\phi \eqendspace.
\ee
This relation actually imposes a constraint on phase space, that has to be treated with some care. We will work in the extended phase space parametrized by canonically conjugate variables $(\phi,\pi)$ and impose the constraint {\em after} evaluating the Poisson brackets as is standard in these cases. We will use the symbol $=$ for results in the extended space and $\approx$ for results after imposing the constraint.

In the Hamiltonian formalism, the corresponding Hamiltonian and momentum densities differ from Eqs. \eqref{eq:chdensities} by terms proportional to the constraint, that vanish when evaluated on the physical constrained phase space
\begin{subequations}
\begin{align}
    &\hdens = e+\lambda_t \left( \pi-\frac{1}{2}\partial_s\phi\right) \eqendspace, \\
    &\momdens_i = k_i+\lambda_i\left( \pi-\frac{1}{2}\partial_s\phi\right) \eqendspace, \\
    &\chdens= n +\lambda\left( \pi-\frac{1}{2}\partial_s\phi\right) \eqendspace.
\end{align}
\end{subequations}
The Lagrange multipliers are fixed by the condition that the charges act as the generators of time and space translations and shifts of the scalar on the physical space. The Hamiltonian, momentum and charge are
\be
H=\int d^2 x\, \hdens \eqendspace,\ \ \mom_i=\int d^2 x \, \momdens_i \eqendspace,\ \ \charge = \int d^2 x\, \chdens \eqendspace.
\ee
Then,
\be
\{ \phi, H\}\approx \lambda_t \eqendspace,\ \ \{ \phi, \mom_i\} \approx \lambda_i \eqendspace,\ \ \{\phi, \charge \}\approx \lambda \eqendspace.
\ee
Which fixes $\lambda_t = \partial_t \phi$, $\lambda_i = -\partial_i \phi$ and $\lambda=1$. In this case the densities $\hdens$, $\momdens_i$ and $\chdens$ in the extended phase space take the usual form for an ordinary scalar, so they lead to the same Poisson brackets and hydrodynamic equations in the extended space. Imposing the contraint fixes the density and the momentum in the direction transverse to the Burgers vector in terms of the same quantity
\be
\chdens \approx \frac{1}{2}\partial_s\phi \eqendspace, \ \ \momdens_s\approx -\frac{1}{2}(\partial_s\phi)^2 \eqendspace.
\ee
This can be interpreted as an equation of state for the fractonic fluid relating the momentum and the disclination density
\be
\momdens_s = -2 \chdens^2 \eqendspace.
\ee
In this case the hydrodynamic equation for $\momdens_s$ becomes a constraint.

Solving for the spatial derivatives of $\phi$ and the conjugate momentum, the Hamiltonian is
\begin{align}
    & \hdens = \momdens_b + \frac{(\lambda_t-\lambda_b)}{4\lambda^2} \Bigl( \lambda_s+4\lambda \chdens \notag \\
    &\hspace{2.3cm} + \operatorname{sign}(\lambda_s)\sqrt{\lambda_s^2+8\lambda(\lambda_s \chdens - \lambda \momdens_s)} \Bigr) \eqendspace.
\end{align}
We will take this as the Hamiltonian for a hydrodynamic theory. For $\lambda_s\neq 0$ one finds that, imposing the conditions,
\be
\frac{\partial \hdens}{\partial \lambda_\mu}=0 \eqendspace,\ \ \frac{\partial \hdens}{\partial \lambda}=0 \eqendspace,
\ee
fixes
\be
\momdens_s = -2 \chdens^2 \eqendspace,\ \ \lambda_s=-2 \lambda \chdens \eqendspace.
\ee
These are the same conditions we have in the microscopic model.

Then, identifying
\be
\mu=\frac{\partial \hdens}{\partial \chdens} \eqendspace,\ \ v^a=\frac{\partial \hdens}{\partial \momdens_a} \eqendspace,
\ee
subject to the previous conditions, one gets
\be
\lambda_t=\lambda_b+\frac{\lambda}{2}\mu \eqendspace,\ \ v^s=\frac{\mu}{4 \chdens},\ \ v^b=1 \eqendspace.
\ee
And, evaluating the Hamiltonian,
\be
\hdens = \momdens_b \eqendspace.
\ee
Thus the energy equals the momentum in the direction of the Burgers vector. Note that we should impose the conditions {\em after} deriving the equations.

The hydrodynamic equation for $\momdens_s$ becomes proportional to the equation for $\chdens$, but is missing a term depending on the chemical potential. We are left with two independent equations and a constraint on $\mu$:
\begin{subequations}
\begin{align}
    & \partial_t{\chdens}+\partial_b \chdens = 0 \eqendspace,\ \ \partial_s \mu=0 \eqendspace, \\
    &\partial_t \momdens_b+\partial_b\left( \momdens_b+\frac{1}{2}\mu \chdens \right)+ \partial_s \left( \momdens_b \frac{\mu}{4 \chdens} \right) =0 \eqendspace.
\end{align}
\end{subequations}
The equation for $\chdens$ coincides with the equation of motion in the microscopic model. Note that $\mu$ is not a function of the density or the momenta, rather its role is to ensure that the symmetry \eqref{eq:symmtranspb} is realized in the hydrodynamic equations, as we will see presently.

Expanding to linear order around constant background values $(\chdens_0,\mu_0)$ for the density and chemical potential, we get the set of equations
\begin{subequations}
\begin{align}
    & \partial_t\delta{\chdens}+\partial_b \delta\chdens = 0 \eqendspace,\ \ \partial_s \delta\mu=0 \eqendspace, \\
    &\partial_t\delta \momdens_b+\partial_b\left(\delta \momdens_b+\frac{1}{2}\mu_0 \delta\chdens+\frac{1}{2}\chdens_0 \delta\mu\right)+ \frac{\mu_0}{4\chdens_0}\partial_s \delta \momdens_b =0 \eqendspace.
\end{align}
\end{subequations}
The solutions for the density and chemical potential should take the form
\be
\delta\chdens = \delta\bar{\chdens}(b-t,s) + \chdens_0 \alpha(b-t) \eqendspace,\ \ \delta \mu=\delta\mu(t,b) \eqendspace,
\ee
where $\partial_s\delta\bar{\chdens}\neq 0$. We have to distinguish between the contributions to the chemical potential that depend on $b-t$ and the remainder. We will split the chemical potential as follows
\be\label{eq:mufluc}
\delta \mu=\mu_0\beta(b-t)+(\partial_t+\partial_b)\varphi(t,b) \eqendspace.
\ee
In the momentum we can separate a part that will give a contribution independent of the coordinate $s$ to the conservation equation from another part that can in principle have an arbitrary dependence on $s$:
\be
\delta \momdens_b = \delta \bar{\momdens}_b(t,b,s) + \chdens_0\left[\psi(t,b)+\chdens_0(s-s_0)\gamma(b-t)\right] \eqendspace.
\ee
We are left with the equations
\begin{subequations}
\begin{align}
    &(\partial_t+\partial_b)\delta\bar{\momdens}_b+\frac{\mu_0}{2}\partial_b\delta\bar{\chdens}+ \frac{\mu_0}{4\chdens_0}\partial_s \delta \bar{\momdens}_b =0 \eqendspace, \\
    &\chdens_0\left(\partial_t+\partial_b \right)\left(\psi+\frac{1}{2}\partial_b\varphi\right) \notag \\
    &\hspace{2.35cm} +\frac{\mu_0\chdens_0}{2}\left(\partial_b\alpha+ \partial_b\beta+\frac{1}{2}\gamma \right)=0 \eqendspace.
\end{align}
\end{subequations}
The solutions are
\begin{subequations} \label{eq:solutionsfrac}
\begin{align}
    &\gamma=-2\partial_b\alpha-2\partial_b\beta \eqendspace, \ \ \psi=-\frac{1}{2}\partial_b \varphi \eqendspace, \\
    &\delta \bar{\chdens}=\partial_s  \delta \bar{J} \eqendspace, \ \ \delta \bar{J}=\delta \bar{J}(b-t,s) \eqendspace, \\
    &\delta \bar{\momdens}_b=-2\chdens_0 \partial_b \delta \bar{J}+\delta \momdens_b^h \eqendspace.
\end{align}
\end{subequations}
The homogeneous term $\delta \momdens_b^h$ can be expanded in plane waves propagating with speed unity in the $b$ direction and speed $c_s=\mu_0/(4 \chdens_0)=v_0^s$ in the $s$ direction
\be\label{eq:transversechiral}
\delta \momdens_b^h=\int \frac{d^2q}{(2\pi)^2} \, \widetilde{\momdens}_b (q) e^{i q_s(s-c_st)+i q_b (b-t)} \eqendspace.
\ee
We can identify $\delta \bar{J}$ as part of the current, with an additional contribution parametrized by $\alpha$
\be
\delta J=\delta \bar{J}+\chdens_0(s-s_0)\alpha \eqendspace.
\ee
The solution for the chemical potential \eqref{eq:mufluc} confirms its part in the realization of the symmetries. Comparing with the symmetry transformation of $\momdens_b$ \eqref{eq:symmtranspb} generated by $g(t,b)$ (the constraint has already been imposed at this point and so $\momdens_b$ and $k_b$ are identified), we see that $\varphi$ and $\psi$ correspond to said transformation, with a condition $(\partial_t+\partial_b)g(t,b)\neq 0$.

Thus, there are three types of modes propagating momentum $\momdens_b$: the ones associated to the current $\delta J$, the ones appearing in the homogeneous contribution $\delta \momdens_b^h$ and the mode $\beta$. This last mode can be absorbed in the homogeneous part when $\mu_0\to 0$ so that $c_s\to 0$. \\


\section{Discussion}

We have derived hydrodynamic equations and constitutive relations for ideal fluids with a dipole symmetry. Our derivation utilizes representative microscopic  scalar models and the Poisson brackets between densities of conserved charges, following the classical approach pioneered by Landau. We have considered three types of models, two that we dub ``non-chiral fracton and a third one ``chiral fracton'' that contains a vector that can be identified with the Burgers vector in the fracton description of disclinations and dislocations. In the former we introduce two spatial derivatives per field, so that there is  either a shift  or U(1) symmetry linear in the spatial coordinates, and in the case of the real field, allow terms with an arbitrary number of fields.  In the chiral fracton model, there is just one derivative per field and is truncated to quadratic order. All models produce the same hydrodynamic equations when written in terms of the densities but with distinct constitutive relations.
 
In all cases the particle number current is a total derivative either by construction in the non-chiral models \eqref{eq_momentumchempot}, or on-shell in the chiral fracton model \eqref{eq:solutionsfrac}, in such a way that the dipole moment of the charge is conserved in both. We have studied the spectrum of fluctuations around a state with fixed monopole density. In the non-chiral case there are modes with a dispersion relation $\omega^2\sim  q^4$ \eqref{eq:dispersionsimple}, where $q^4$ stands for a general quartic polynomial of the momenta, with coefficients determined by derivatives of the energy density with respect to the gradient of the generalized velocity $V_i=\chdens^{-1}\momdens_i$. In the chiral fracton model there are chiral modes propagating in the direction of the Burgers vector as well as a mode that also propagates energy and momentum in the transverse direction \eqref{eq:transversechiral} with a velocity propotional to the chemical potential.

In \cite{GromovLucas2020} charge difussion was studied for a model sitting within the non-chiral systems we considered. However, they did not consider momentum conservation. A complete finite-temperature hydrodynamic theory should account for both charge and momentum conservation, as we have done in this paper. But, the theory should also include dissipative effects, which is an aspect that we leave for future studies. Nonetheless, a peculiarity of systems with MDMA symmetry is the presence of a non-ballistic propagating mode and a diffusive one with dispersion relation $\omega\sim -iq^4$ \cite{GromovLucas2020}, contrary to ordinary hydrodynamics where sound waves are ballistic and have a quadratic dispersion relation for diffusive modes.

On the other hand, defect dynamics has important ramifications in crystals, amorphous solids, liquid crystals and active matter \cite{Kleman2008review,Shankar2020}. Although the continuous description of defects as a gauge theory has a long history, a symmetry-based construction of defect hydrodynamics beyond vortices is still lacking and various phenomenological equations have been proposed for specific systems (see e.g. \cite{Mura1963,Kosevich_1965,Julia1979,Acharya2001,FRESSENGEAS20113499,Toth2002,Tang2019,Acharya2021}). In order to fill this gap, we have proposed a Poisson bracket derivation of hydrodynamic equations given two main ingredients that are important in modelling macroscopic systems of defects. The first ingredient is invariance under MDMA symmetry, the second ingredient is the existence of a Burgers vector. These requirements can be viewed as basic guiding principles for hydrodynamic theories of defects as can be justified using the dual fracton picture. A model displaying the MDMA symmetry and a Burgers vector is the chiral fracton theory of a scalar field. Interpreting the model as a description of lattice defects, the chiral propagation of the disclination density has a simple visualization. In any symmetric region of finite area there is a finite number of disclinations determined by the integration of the constant charge density in the region. The dipole moment on the other hand vanishes, indicating a net zero charge of dislocations. However, as the dislocation density is not identically zero, there is a nonzero amount of dislocation pairs. While immobile in isolation, disclinations can move by absorbing or emitting dislocation pairs \cite{Harris:1971,deWit:1971,GUTKIN200373,desai_kapral_2009,FRESSENGEAS20113499}, in which case their motion is along the Burgers vector of the dislocations. This matches nicely with the behavior of the solutions we obtain in the chiral fracton model. Thus the results of this paper provide basic building blocks to construct more realistic models and ultimately study transport properties of defects.


\emph{Acknowledgements.} K.T.G., P.S.  and F.P.-B. acknowledge financial support by
the Deutsche Forschungsgemeinschaft (DFG, German Research Foundation) under Germany’s Excellence Strategy
through the Würzburg-Dresden Cluster of Excellence on Complexity and Topology in Quantum Matter - ct.qmat
(EXC 2147, project-id 390858490). K.T.G. also acknowledges the support of the Hallwachs-Röntgen Postdoc Program of ct.qmat. C.H. has been partially supported by the Spanish {\em Ministerio de Ciencia, Innovación y Universidades} through Grant PGC2018-096894-B-100. P.S. was supported by the Deutsche Forschungsgemeinschaft through
the Leibniz Program, and the Narodowe Centrum Nauki (Polish National Science Centre) Sonata Bis Grant
2019/34/E/ST3/00405.


\appendix

\section{More general Lagrangians}

Suppose we allow for the more general dependence of the Lagrangian density \eqref{eq:L} of the form
\begin{equation} \label{eq:Lgen}
	L=\int\vol\,\mathcal L(\partial_t\phi, \partial_i \phi , \partial_i\partial_j\phi , \ldots )\eqendspace,
\end{equation}
where $\ldots$ stands for higher spatial derivatives of $\phi$. Let us define the variational quantities
\begin{subequations}
\begin{align}
    \pi &= \frac{\partial \mathcal{L}}{\partial ( \partial_t \phi )} \eqendspace, \\
    \pi^{i_1 \cdots i_n} &= (-1)^{n-1} \frac{\partial \mathcal{L}}{\partial ( \partial_{i_1} \cdots \partial_{i_n} \phi )}\,, \quad n \geq 1\eqendspace.
\end{align}
\end{subequations}
For simplicity we introduce the notation  $\veci_n = (i_1 , \ldots , i_n )$ to be an $n$-tuple of indices and
\begin{align}
    \pi^{\veci_n} &= \pi^{i_1 \cdots i_n}\eqendspace, &%
    \partial_{\veci_n}^{n} &= \partial_{i_1} \cdots \partial_{i_n}\eqendspace.
\end{align}
Indices can be concatenated, $\pi^{i \vecj_n} = \pi^{ij_1 \cdots j_n}$, and similarly for the partial derivatives, $\partial_{i \vecj_n}^{n+1} = \partial_i \partial_{j_1} \cdots \partial_{j_n}$.

Then, as discussed earlier, for this theory to be invariant under the linear shift, we require that
\begin{equation}
    \pi^i = \frac{\partial \mathcal{L}}{\partial ( \partial_i \phi )} = \partial^i \chi \eqendspace,
\end{equation}
for some functional $\chi$ of $\phi$ and its derivatives. 

With this in place, the only difference between this case and the simpler case considered in the main text is that some expressions become more complicated. The density $\chdens$ is still defined to be $\chdens = \pi$. However, the expression for $J^{ij}$ in \eqref{eq:piandJ} is modified to
\begin{align}
   J^{ij} = \chi \delta^{ij} + \sum_{n \geq 0} \partial_{\vec{k}_n}^{n} \pi^{ij \vec{k}_n}\eqendspace.
\end{align}
This reduces to the simpler case studied in the main text when we set $\chi = 0$ and $\pi^{\veci_n} = 0$ for $n \geq 3$.

The Hamiltonian and momentum densities are unchanged,
\begin{subequations}
\begin{align}
    \hdens &= \pi \, \partial_t \phi - \mathcal{L} \eqendspace, \\
    \momdens_i &= -\pi \, \partial_i \phi \eqendspace,
\end{align}
\end{subequations}
but the corresponding spatial currents \eqref{eq:Tit} and \eqref{eq:Tji} are modified to
\begin{subequations}
\begin{align}
    \EMT^{i}{}_{t} &= ( \partial_j J^{ij} ) \partial_t \phi + ( \chi \delta^{ij} - J^{ij} ) \partial_j \partial_t \phi \notag \\
    &\quad + \sum_{m \geq 2} \sum_{n \geq 0} (-1)^m \partial_{\vec{k}_n}^n \pi^{i \vecj_m \vec{k}_n} \partial_{\vecj_m} \partial_t \phi\eqendspace, \\
    \EMT^{j}{}_{i} &= - ( \partial_k J^{jk} ) \partial_i \phi + ( J^{jk} - \chi \delta^{jk} ) \partial_k \partial_i \phi + \delta^{j}_{i} \mathcal{L} \notag \\
    &\quad - \sum_{m \geq 2} \sum_{n \geq 0} (-1)^m \partial_{\vec{k}_n}^{n} \pi^{j \vec{\ell}_m \vec{k}_n} \partial_{\vec{\ell}_m}^{m} \partial_i \phi\eqendspace.
\end{align}
\end{subequations}
The equation of motion \eqref{eq:eom}, continuity equations \eqref{eq:ctyeqs} and current relations \eqref{eq:shiftjrels} associated with the constant and linear shift symmetries, and the energy-momentum conservation equations \eqref{eq:EMcons} associated with time- and space-translation invariance all still hold in precisely the same form.


\begin{thebibliography}{80}%
	\makeatletter
	\providecommand \@ifxundefined [1]{%
		\@ifx{#1\undefined}
	}%
	\providecommand \@ifnum [1]{%
		\ifnum #1\expandafter \@firstoftwo
		\else \expandafter \@secondoftwo
		\fi
	}%
	\providecommand \@ifx [1]{%
		\ifx #1\expandafter \@firstoftwo
		\else \expandafter \@secondoftwo
		\fi
	}%
	\providecommand \natexlab [1]{#1}%
	\providecommand \enquote  [1]{``#1''}%
	\providecommand \bibnamefont  [1]{#1}%
	\providecommand \bibfnamefont [1]{#1}%
	\providecommand \citenamefont [1]{#1}%
	\providecommand \href@noop [0]{\@secondoftwo}%
	\providecommand \href [0]{\begingroup \@sanitize@url \@href}%
	\providecommand \@href[1]{\@@startlink{#1}\@@href}%
	\providecommand \@@href[1]{\endgroup#1\@@endlink}%
	\providecommand \@sanitize@url [0]{\catcode `\\12\catcode `\$12\catcode
		`\&12\catcode `\#12\catcode `\^12\catcode `\_12\catcode `\%12\relax}%
	\providecommand \@@startlink[1]{}%
	\providecommand \@@endlink[0]{}%
	\providecommand \url  [0]{\begingroup\@sanitize@url \@url }%
	\providecommand \@url [1]{\endgroup\@href {#1}{\urlprefix }}%
	\providecommand \urlprefix  [0]{URL }%
	\providecommand \Eprint [0]{\href }%
	\providecommand \doibase [0]{https://doi.org/}%
	\providecommand \selectlanguage [0]{\@gobble}%
	\providecommand \bibinfo  [0]{\@secondoftwo}%
	\providecommand \bibfield  [0]{\@secondoftwo}%
	\providecommand \translation [1]{[#1]}%
	\providecommand \BibitemOpen [0]{}%
	\providecommand \bibitemStop [0]{}%
	\providecommand \bibitemNoStop [0]{.\EOS\space}%
	\providecommand \EOS [0]{\spacefactor3000\relax}%
	\providecommand \BibitemShut  [1]{\csname bibitem#1\endcsname}%
	\let\auto@bib@innerbib\@empty
	\bibitem [{\citenamefont {Nandkishore}\ and\ \citenamefont
		{Hermele}(2019)}]{nandkishore_fractons_2019}%
	\BibitemOpen
	\bibfield  {author} {\bibinfo {author} {\bibfnamefont {R.~M.}\ \bibnamefont
			{Nandkishore}}\ and\ \bibinfo {author} {\bibfnamefont {M.}~\bibnamefont
			{Hermele}},\ }\bibfield  {title} {\bibinfo {title} {Fractons},\ }\href
	{https://doi.org/10.1146/annurev-conmatphys-031218-013604} {\bibfield
		{journal} {\bibinfo  {journal} {Annual Review of Condensed Matter Physics}\
		}\textbf {\bibinfo {volume} {10}},\ \bibinfo {pages} {295} (\bibinfo {year}
		{2019})}\BibitemShut {NoStop}%
	\bibitem [{\citenamefont {Pretko}\ \emph {et~al.}(2020)\citenamefont {Pretko},
		\citenamefont {Chen},\ and\ \citenamefont {You}}]{Pretko:2020cko}%
	\BibitemOpen
	\bibfield  {author} {\bibinfo {author} {\bibfnamefont {M.}~\bibnamefont
			{Pretko}}, \bibinfo {author} {\bibfnamefont {X.}~\bibnamefont {Chen}},\ and\
		\bibinfo {author} {\bibfnamefont {Y.}~\bibnamefont {You}},\ }\bibfield
	{title} {\bibinfo {title} {{Fracton Phases of Matter}},\ }\href
	{https://doi.org/10.1142/S0217751X20300033} {\bibfield  {journal} {\bibinfo
			{journal} {Int. J. Mod. Phys. A}\ }\textbf {\bibinfo {volume} {35}},\
		\bibinfo {pages} {2030003} (\bibinfo {year} {2020})}\BibitemShut {NoStop}%
	\bibitem [{\citenamefont {Chamon}(2005)}]{Chamon:2005}%
	\BibitemOpen
	\bibfield  {author} {\bibinfo {author} {\bibfnamefont {C.}~\bibnamefont
			{Chamon}},\ }\bibfield  {title} {\bibinfo {title} {Quantum glassiness in
			strongly correlated clean systems: An example of topological
			overprotection},\ }\href {https://doi.org/10.1103/PhysRevLett.94.040402}
	{\bibfield  {journal} {\bibinfo  {journal} {Phys. Rev. Lett.}\ }\textbf
		{\bibinfo {volume} {94}},\ \bibinfo {pages} {040402} (\bibinfo {year}
		{2005})}\BibitemShut {NoStop}%
	\bibitem [{\citenamefont {Haah}(2011)}]{Haah2011}%
	\BibitemOpen
	\bibfield  {author} {\bibinfo {author} {\bibfnamefont {J.}~\bibnamefont
			{Haah}},\ }\bibfield  {title} {\bibinfo {title} {Local stabilizer codes in
			three dimensions without string logical operators},\ }\href
	{https://doi.org/10.1103/PhysRevA.83.042330} {\bibfield  {journal} {\bibinfo
			{journal} {Phys. Rev. A}\ }\textbf {\bibinfo {volume} {83}},\ \bibinfo
		{pages} {042330} (\bibinfo {year} {2011})}\BibitemShut {NoStop}%
	\bibitem [{\citenamefont {Bravyi}\ \emph {et~al.}(2011)\citenamefont {Bravyi},
		\citenamefont {Leemhuis},\ and\ \citenamefont {Terhal}}]{Bravyi:2011}%
	\BibitemOpen
	\bibfield  {author} {\bibinfo {author} {\bibfnamefont {S.}~\bibnamefont
			{Bravyi}}, \bibinfo {author} {\bibfnamefont {B.}~\bibnamefont {Leemhuis}},\
		and\ \bibinfo {author} {\bibfnamefont {B.~M.}\ \bibnamefont {Terhal}},\
	}\bibfield  {title} {\bibinfo {title} {Topological order in an exactly
			solvable 3d spin model},\ }\href
	{https://doi.org/https://doi.org/10.1016/j.aop.2010.11.002} {\bibfield
		{journal} {\bibinfo  {journal} {Annals of Physics}\ }\textbf {\bibinfo
			{volume} {326}},\ \bibinfo {pages} {839} (\bibinfo {year}
		{2011})}\BibitemShut {NoStop}%
	\bibitem [{\citenamefont {Vijay}\ \emph {et~al.}(2016)\citenamefont {Vijay},
		\citenamefont {Haah},\ and\ \citenamefont {Fu}}]{Vijay:2016phm}%
	\BibitemOpen
	\bibfield  {author} {\bibinfo {author} {\bibfnamefont {S.}~\bibnamefont
			{Vijay}}, \bibinfo {author} {\bibfnamefont {J.}~\bibnamefont {Haah}},\ and\
		\bibinfo {author} {\bibfnamefont {L.}~\bibnamefont {Fu}},\ }\bibfield
	{title} {\bibinfo {title} {{Fracton Topological Order, Generalized Lattice
				Gauge Theory and Duality}},\ }\href
	{https://doi.org/10.1103/PhysRevB.94.235157} {\bibfield  {journal} {\bibinfo
			{journal} {Phys. Rev. B}\ }\textbf {\bibinfo {volume} {94}},\ \bibinfo
		{pages} {235157} (\bibinfo {year} {2016})}\BibitemShut {NoStop}%
	\bibitem [{\citenamefont {Williamson}(2016)}]{Williamson2016}%
	\BibitemOpen
	\bibfield  {author} {\bibinfo {author} {\bibfnamefont {D.~J.}\ \bibnamefont
			{Williamson}},\ }\bibfield  {title} {\bibinfo {title} {Fractal symmetries:
			Ungauging the cubic code},\ }\href
	{https://doi.org/10.1103/PhysRevB.94.155128} {\bibfield  {journal} {\bibinfo
			{journal} {Phys. Rev. B}\ }\textbf {\bibinfo {volume} {94}},\ \bibinfo
		{pages} {155128} (\bibinfo {year} {2016})}\BibitemShut {NoStop}%
	\bibitem [{\citenamefont {Xu}(2006)}]{Xu2006}%
	\BibitemOpen
	\bibfield  {author} {\bibinfo {author} {\bibfnamefont {C.}~\bibnamefont
			{Xu}},\ }\bibfield  {title} {\bibinfo {title} {Gapless bosonic excitation
			without symmetry breaking: An algebraic spin liquid with soft gravitons},\
	}\href {https://doi.org/10.1103/PhysRevB.74.224433} {\bibfield  {journal}
		{\bibinfo  {journal} {Phys. Rev. B}\ }\textbf {\bibinfo {volume} {74}},\
		\bibinfo {pages} {224433} (\bibinfo {year} {2006})}\BibitemShut {NoStop}%
	\bibitem [{\citenamefont {Xu}\ and\ \citenamefont {Ho\ifmmode~\check{r}\else
			\v{r}\fi{}ava}(2010)}]{Xu2010}%
	\BibitemOpen
	\bibfield  {author} {\bibinfo {author} {\bibfnamefont {C.}~\bibnamefont
			{Xu}}\ and\ \bibinfo {author} {\bibfnamefont {P.}~\bibnamefont
			{Ho\ifmmode~\check{r}\else \v{r}\fi{}ava}},\ }\bibfield  {title} {\bibinfo
		{title} {Emergent gravity at a lifshitz point from a bose liquid on the
			lattice},\ }\href {https://doi.org/10.1103/PhysRevD.81.104033} {\bibfield
		{journal} {\bibinfo  {journal} {Phys. Rev. D}\ }\textbf {\bibinfo {volume}
			{81}},\ \bibinfo {pages} {104033} (\bibinfo {year} {2010})}\BibitemShut
	{NoStop}%
	\bibitem [{\citenamefont {Pretko}(2017{\natexlab{a}})}]{Pretko2017spinliquid}%
	\BibitemOpen
	\bibfield  {author} {\bibinfo {author} {\bibfnamefont {M.}~\bibnamefont
			{Pretko}},\ }\bibfield  {title} {\bibinfo {title} {Subdimensional particle
			structure of higher rank $u(1)$ spin liquids},\ }\href
	{https://doi.org/10.1103/PhysRevB.95.115139} {\bibfield  {journal} {\bibinfo
			{journal} {Phys. Rev. B}\ }\textbf {\bibinfo {volume} {95}},\ \bibinfo
		{pages} {115139} (\bibinfo {year} {2017}{\natexlab{a}})}\BibitemShut
	{NoStop}%
	\bibitem [{\citenamefont {Pretko}(2017{\natexlab{b}})}]{Pretko2017spinliquid2}%
	\BibitemOpen
	\bibfield  {author} {\bibinfo {author} {\bibfnamefont {M.}~\bibnamefont
			{Pretko}},\ }\bibfield  {title} {\bibinfo {title} {Generalized
			electromagnetism of subdimensional particles: A spin liquid story},\ }\href
	{https://doi.org/10.1103/PhysRevB.96.035119} {\bibfield  {journal} {\bibinfo
			{journal} {Phys. Rev. B}\ }\textbf {\bibinfo {volume} {96}},\ \bibinfo
		{pages} {035119} (\bibinfo {year} {2017}{\natexlab{b}})}\BibitemShut
	{NoStop}%
	\bibitem [{\citenamefont {You}\ \emph {et~al.}(2020)\citenamefont {You},
		\citenamefont {Bi},\ and\ \citenamefont {Pretko}}]{You_emergent_2020}%
	\BibitemOpen
	\bibfield  {author} {\bibinfo {author} {\bibfnamefont {Y.}~\bibnamefont
			{You}}, \bibinfo {author} {\bibfnamefont {Z.}~\bibnamefont {Bi}},\ and\
		\bibinfo {author} {\bibfnamefont {M.}~\bibnamefont {Pretko}},\ }\bibfield
	{title} {\bibinfo {title} {Emergent fractons and algebraic quantum liquid
			from plaquette melting transitions},\ }\href
	{https://doi.org/10.1103/PhysRevResearch.2.013162} {\bibfield  {journal}
		{\bibinfo  {journal} {Phys. Rev. Research}\ }\textbf {\bibinfo {volume}
			{2}},\ \bibinfo {pages} {013162} (\bibinfo {year} {2020})}\BibitemShut
	{NoStop}%
	\bibitem [{\citenamefont {Pai}\ \emph {et~al.}(2019)\citenamefont {Pai},
		\citenamefont {Pretko},\ and\ \citenamefont {Nandkishore}}]{Pai2019}%
	\BibitemOpen
	\bibfield  {author} {\bibinfo {author} {\bibfnamefont {S.}~\bibnamefont
			{Pai}}, \bibinfo {author} {\bibfnamefont {M.}~\bibnamefont {Pretko}},\ and\
		\bibinfo {author} {\bibfnamefont {R.~M.}\ \bibnamefont {Nandkishore}},\
	}\bibfield  {title} {\bibinfo {title} {Localization in fractonic random
			circuits},\ }\href {https://doi.org/10.1103/PhysRevX.9.021003} {\bibfield
		{journal} {\bibinfo  {journal} {Phys. Rev. X}\ }\textbf {\bibinfo {volume}
			{9}},\ \bibinfo {pages} {021003} (\bibinfo {year} {2019})}\BibitemShut
	{NoStop}%
	\bibitem [{\citenamefont {Feldmeier}\ \emph {et~al.}(2020)\citenamefont
		{Feldmeier}, \citenamefont {Sala}, \citenamefont {De~Tomasi}, \citenamefont
		{Pollmann},\ and\ \citenamefont {Knap}}]{Feldmeier2020}%
	\BibitemOpen
	\bibfield  {author} {\bibinfo {author} {\bibfnamefont {J.}~\bibnamefont
			{Feldmeier}}, \bibinfo {author} {\bibfnamefont {P.}~\bibnamefont {Sala}},
		\bibinfo {author} {\bibfnamefont {G.}~\bibnamefont {De~Tomasi}}, \bibinfo
		{author} {\bibfnamefont {F.}~\bibnamefont {Pollmann}},\ and\ \bibinfo
		{author} {\bibfnamefont {M.}~\bibnamefont {Knap}},\ }\bibfield  {title}
	{\bibinfo {title} {Anomalous diffusion in dipole- and
			higher-moment-conserving systems},\ }\href
	{https://doi.org/10.1103/PhysRevLett.125.245303} {\bibfield  {journal}
		{\bibinfo  {journal} {Phys. Rev. Lett.}\ }\textbf {\bibinfo {volume} {125}},\
		\bibinfo {pages} {245303} (\bibinfo {year} {2020})}\BibitemShut {NoStop}%
	\bibitem [{\citenamefont {Morningstar}\ \emph {et~al.}(2020)\citenamefont
		{Morningstar}, \citenamefont {Khemani},\ and\ \citenamefont
		{Huse}}]{Morningstar2020}%
	\BibitemOpen
	\bibfield  {author} {\bibinfo {author} {\bibfnamefont {A.}~\bibnamefont
			{Morningstar}}, \bibinfo {author} {\bibfnamefont {V.}~\bibnamefont
			{Khemani}},\ and\ \bibinfo {author} {\bibfnamefont {D.~A.}\ \bibnamefont
			{Huse}},\ }\bibfield  {title} {\bibinfo {title} {Kinetically constrained
			freezing transition in a dipole-conserving system},\ }\href
	{https://doi.org/10.1103/PhysRevB.101.214205} {\bibfield  {journal} {\bibinfo
			{journal} {Phys. Rev. B}\ }\textbf {\bibinfo {volume} {101}},\ \bibinfo
		{pages} {214205} (\bibinfo {year} {2020})}\BibitemShut {NoStop}%
	\bibitem [{\citenamefont {Iaconis}\ \emph {et~al.}(2021)\citenamefont
		{Iaconis}, \citenamefont {Lucas},\ and\ \citenamefont
		{Nandkishore}}]{Iaconis2021}%
	\BibitemOpen
	\bibfield  {author} {\bibinfo {author} {\bibfnamefont {J.}~\bibnamefont
			{Iaconis}}, \bibinfo {author} {\bibfnamefont {A.}~\bibnamefont {Lucas}},\
		and\ \bibinfo {author} {\bibfnamefont {R.}~\bibnamefont {Nandkishore}},\
	}\bibfield  {title} {\bibinfo {title} {Multipole conservation laws and
			subdiffusion in any dimension},\ }\href
	{https://doi.org/10.1103/PhysRevE.103.022142} {\bibfield  {journal} {\bibinfo
			{journal} {Phys. Rev. E}\ }\textbf {\bibinfo {volume} {103}},\ \bibinfo
		{pages} {022142} (\bibinfo {year} {2021})}\BibitemShut {NoStop}%
	\bibitem [{\citenamefont {Moudgalya}\ \emph {et~al.}(2021)\citenamefont
		{Moudgalya}, \citenamefont {Prem}, \citenamefont {Huse},\ and\ \citenamefont
		{Chan}}]{moudgalya2021spectral}%
	\BibitemOpen
	\bibfield  {author} {\bibinfo {author} {\bibfnamefont {S.}~\bibnamefont
			{Moudgalya}}, \bibinfo {author} {\bibfnamefont {A.}~\bibnamefont {Prem}},
		\bibinfo {author} {\bibfnamefont {D.~A.}\ \bibnamefont {Huse}},\ and\
		\bibinfo {author} {\bibfnamefont {A.}~\bibnamefont {Chan}},\ }\href@noop {}
	{\bibinfo {title} {Spectral statistics in constrained many-body quantum
			chaotic systems}} (\bibinfo {year} {2021}),\ \Eprint
	{https://arxiv.org/abs/2009.11863} {arXiv:2009.11863 [cond-mat.stat-mech]}
	\BibitemShut {NoStop}%
	\bibitem [{\citenamefont {Pretko}\ and\ \citenamefont
		{Radzihovsky}(2018)}]{PretkoSolid2018}%
	\BibitemOpen
	\bibfield  {author} {\bibinfo {author} {\bibfnamefont {M.}~\bibnamefont
			{Pretko}}\ and\ \bibinfo {author} {\bibfnamefont {L.}~\bibnamefont
			{Radzihovsky}},\ }\bibfield  {title} {\bibinfo {title} {Symmetry-enriched
			fracton phases from supersolid duality},\ }\href
	{https://doi.org/10.1103/PhysRevLett.121.235301} {\bibfield  {journal}
		{\bibinfo  {journal} {Phys. Rev. Lett.}\ }\textbf {\bibinfo {volume} {121}},\
		\bibinfo {pages} {235301} (\bibinfo {year} {2018})}\BibitemShut {NoStop}%
	\bibitem [{\citenamefont {Gromov}(2019{\natexlab{a}})}]{Gromov2019elastic}%
	\BibitemOpen
	\bibfield  {author} {\bibinfo {author} {\bibfnamefont {A.}~\bibnamefont
			{Gromov}},\ }\bibfield  {title} {\bibinfo {title} {Chiral topological
			elasticity and fracton order},\ }\href
	{https://doi.org/10.1103/PhysRevLett.122.076403} {\bibfield  {journal}
		{\bibinfo  {journal} {Phys. Rev. Lett.}\ }\textbf {\bibinfo {volume} {122}},\
		\bibinfo {pages} {076403} (\bibinfo {year} {2019}{\natexlab{a}})}\BibitemShut
	{NoStop}%
	\bibitem [{\citenamefont {Kumar}\ and\ \citenamefont
		{Potter}(2019)}]{Kumar2019}%
	\BibitemOpen
	\bibfield  {author} {\bibinfo {author} {\bibfnamefont {A.}~\bibnamefont
			{Kumar}}\ and\ \bibinfo {author} {\bibfnamefont {A.~C.}\ \bibnamefont
			{Potter}},\ }\bibfield  {title} {\bibinfo {title} {Symmetry-enforced
			fractonicity and two-dimensional quantum crystal melting},\ }\href
	{https://doi.org/10.1103/PhysRevB.100.045119} {\bibfield  {journal} {\bibinfo
			{journal} {Phys. Rev. B}\ }\textbf {\bibinfo {volume} {100}},\ \bibinfo
		{pages} {045119} (\bibinfo {year} {2019})}\BibitemShut {NoStop}%
	\bibitem [{\citenamefont {{Pretko}}\ and\ \citenamefont
		{{Radzihovsky}}(2018)}]{Pretko:2018fed}%
	\BibitemOpen
	\bibfield  {author} {\bibinfo {author} {\bibfnamefont {M.}~\bibnamefont
			{{Pretko}}}\ and\ \bibinfo {author} {\bibfnamefont {L.}~\bibnamefont
			{{Radzihovsky}}},\ }\bibfield  {title} {\bibinfo {title} {{Fracton-Elasticity
				Duality}},\ }\href {https://doi.org/10.1103/PhysRevLett.120.195301}
	{\bibfield  {journal} {\bibinfo  {journal} {\prl}\ }\textbf {\bibinfo
			{volume} {120}},\ \bibinfo {eid} {195301} (\bibinfo {year}
		{2018})}\BibitemShut {NoStop}%
	\bibitem [{\citenamefont {Pretko}\ \emph {et~al.}(2019)\citenamefont {Pretko},
		\citenamefont {Zhai},\ and\ \citenamefont {Radzihovsky}}]{pretko2019crystal}%
	\BibitemOpen
	\bibfield  {author} {\bibinfo {author} {\bibfnamefont {M.}~\bibnamefont
			{Pretko}}, \bibinfo {author} {\bibfnamefont {Z.}~\bibnamefont {Zhai}},\ and\
		\bibinfo {author} {\bibfnamefont {L.}~\bibnamefont {Radzihovsky}},\
	}\bibfield  {title} {\bibinfo {title} {Crystal-to-fracton tensor gauge theory
			dualities},\ }\href {https://doi.org/10.1103/PhysRevB.100.134113} {\bibfield
		{journal} {\bibinfo  {journal} {Phys. Rev. B}\ }\textbf {\bibinfo {volume}
			{100}},\ \bibinfo {pages} {134113} (\bibinfo {year} {2019})}\BibitemShut
	{NoStop}%
	\bibitem [{\citenamefont {Zhai}\ and\ \citenamefont
		{Radzihovsky}(2019)}]{zhai2019two}%
	\BibitemOpen
	\bibfield  {author} {\bibinfo {author} {\bibfnamefont {Z.}~\bibnamefont
			{Zhai}}\ and\ \bibinfo {author} {\bibfnamefont {L.}~\bibnamefont
			{Radzihovsky}},\ }\bibfield  {title} {\bibinfo {title} {Two-dimensional
			melting via sine-gordon duality},\ }\href
	{https://doi.org/10.1103/PhysRevB.100.094105} {\bibfield  {journal} {\bibinfo
			{journal} {Phys. Rev. B}\ }\textbf {\bibinfo {volume} {100}},\ \bibinfo
		{pages} {094105} (\bibinfo {year} {2019})}\BibitemShut {NoStop}%
	\bibitem [{\citenamefont {Gromov}\ and\ \citenamefont
		{Sur\'owka}(2020)}]{Gromov:2019waa}%
	\BibitemOpen
	\bibfield  {author} {\bibinfo {author} {\bibfnamefont {A.}~\bibnamefont
			{Gromov}}\ and\ \bibinfo {author} {\bibfnamefont {P.}~\bibnamefont
			{Sur\'owka}},\ }\bibfield  {title} {\bibinfo {title} {{On duality between
				Cosserat elasticity and fractons}},\ }\href
	{https://doi.org/10.21468/SciPostPhys.8.4.065} {\bibfield  {journal}
		{\bibinfo  {journal} {SciPost Phys.}\ }\textbf {\bibinfo {volume} {8}},\
		\bibinfo {pages} {065} (\bibinfo {year} {2020})}\BibitemShut {NoStop}%
	\bibitem [{\citenamefont {Nguyen}\ \emph {et~al.}(2020)\citenamefont {Nguyen},
		\citenamefont {Gromov},\ and\ \citenamefont {Moroz}}]{Nguyen:2020yve}%
	\BibitemOpen
	\bibfield  {author} {\bibinfo {author} {\bibfnamefont {D.~X.}\ \bibnamefont
			{Nguyen}}, \bibinfo {author} {\bibfnamefont {A.}~\bibnamefont {Gromov}},\
		and\ \bibinfo {author} {\bibfnamefont {S.}~\bibnamefont {Moroz}},\ }\bibfield
	{title} {\bibinfo {title} {{Fracton-elasticity duality of two-dimensional
				superfluid vortex crystals: defect interactions and quantum melting}},\
	}\href {https://doi.org/10.21468/SciPostPhys.9.5.076} {\bibfield  {journal}
		{\bibinfo  {journal} {SciPost Phys.}\ }\textbf {\bibinfo {volume} {9}},\
		\bibinfo {pages} {076} (\bibinfo {year} {2020})}\BibitemShut {NoStop}%
	\bibitem [{\citenamefont {Fruchart}\ and\ \citenamefont
		{Vitelli}(2020)}]{Fruchart2020dual}%
	\BibitemOpen
	\bibfield  {author} {\bibinfo {author} {\bibfnamefont {M.}~\bibnamefont
			{Fruchart}}\ and\ \bibinfo {author} {\bibfnamefont {V.}~\bibnamefont
			{Vitelli}},\ }\bibfield  {title} {\bibinfo {title} {Symmetries and dualities
			in the theory of elasticity},\ }\href
	{https://doi.org/10.1103/PhysRevLett.124.248001} {\bibfield  {journal}
		{\bibinfo  {journal} {Phys. Rev. Lett.}\ }\textbf {\bibinfo {volume} {124}},\
		\bibinfo {pages} {248001} (\bibinfo {year} {2020})}\BibitemShut {NoStop}%
	\bibitem [{\citenamefont {Manoj}\ \emph {et~al.}(2020)\citenamefont {Manoj},
		\citenamefont {Moessner},\ and\ \citenamefont {Shenoy}}]{Manoj:2020abe}%
	\BibitemOpen
	\bibfield  {author} {\bibinfo {author} {\bibfnamefont {N.}~\bibnamefont
			{Manoj}}, \bibinfo {author} {\bibfnamefont {R.}~\bibnamefont {Moessner}},\
		and\ \bibinfo {author} {\bibfnamefont {V.~B.}\ \bibnamefont {Shenoy}},\
	}\bibfield  {title} {\bibinfo {title} {{Tearing Fractons}},\ }\href@noop {}
	{\  (\bibinfo {year} {2020})},\ \Eprint {https://arxiv.org/abs/2011.11401}
	{arXiv:2011.11401 [cond-mat.str-el]} \BibitemShut {NoStop}%
	\bibitem [{\citenamefont {Sur\'owka}(2021)}]{Surowka:2021ved}%
	\BibitemOpen
	\bibfield  {author} {\bibinfo {author} {\bibfnamefont {P.}~\bibnamefont
			{Sur\'owka}},\ }\bibfield  {title} {\bibinfo {title} {{Dual gauge theory
				formulation of planar quasicrystal elasticity and fractons}},\ }\href@noop {}
	{\  (\bibinfo {year} {2021})},\ \Eprint {https://arxiv.org/abs/2101.12234}
	{arXiv:2101.12234 [cond-mat.str-el]} \BibitemShut {NoStop}%
	\bibitem [{\citenamefont {Paramekanti}\ \emph {et~al.}(2002)\citenamefont
		{Paramekanti}, \citenamefont {Balents},\ and\ \citenamefont
		{Fisher}}]{Paramekanti2002}%
	\BibitemOpen
	\bibfield  {author} {\bibinfo {author} {\bibfnamefont {A.}~\bibnamefont
			{Paramekanti}}, \bibinfo {author} {\bibfnamefont {L.}~\bibnamefont
			{Balents}},\ and\ \bibinfo {author} {\bibfnamefont {M.~P.~A.}\ \bibnamefont
			{Fisher}},\ }\bibfield  {title} {\bibinfo {title} {Ring exchange, the exciton
			bose liquid, and bosonization in two dimensions},\ }\href
	{https://doi.org/10.1103/PhysRevB.66.054526} {\bibfield  {journal} {\bibinfo
			{journal} {Phys. Rev. B}\ }\textbf {\bibinfo {volume} {66}},\ \bibinfo
		{pages} {054526} (\bibinfo {year} {2002})}\BibitemShut {NoStop}%
	\bibitem [{\citenamefont {Sandvik}\ \emph {et~al.}(2002)\citenamefont
		{Sandvik}, \citenamefont {Daul}, \citenamefont {Singh},\ and\ \citenamefont
		{Scalapino}}]{Sandvik2002}%
	\BibitemOpen
	\bibfield  {author} {\bibinfo {author} {\bibfnamefont {A.~W.}\ \bibnamefont
			{Sandvik}}, \bibinfo {author} {\bibfnamefont {S.}~\bibnamefont {Daul}},
		\bibinfo {author} {\bibfnamefont {R.~R.~P.}\ \bibnamefont {Singh}},\ and\
		\bibinfo {author} {\bibfnamefont {D.~J.}\ \bibnamefont {Scalapino}},\
	}\bibfield  {title} {\bibinfo {title} {Striped phase in a quantum $xy$ model
			with ring exchange},\ }\href {https://doi.org/10.1103/PhysRevLett.89.247201}
	{\bibfield  {journal} {\bibinfo  {journal} {Phys. Rev. Lett.}\ }\textbf
		{\bibinfo {volume} {89}},\ \bibinfo {pages} {247201} (\bibinfo {year}
		{2002})}\BibitemShut {NoStop}%
	\bibitem [{\citenamefont {Rousseau}\ \emph {et~al.}(2004)\citenamefont
		{Rousseau}, \citenamefont {Batrouni},\ and\ \citenamefont
		{Scalettar}}]{Rousseau2004}%
	\BibitemOpen
	\bibfield  {author} {\bibinfo {author} {\bibfnamefont {V.}~\bibnamefont
			{Rousseau}}, \bibinfo {author} {\bibfnamefont {G.~G.}\ \bibnamefont
			{Batrouni}},\ and\ \bibinfo {author} {\bibfnamefont {R.~T.}\ \bibnamefont
			{Scalettar}},\ }\bibfield  {title} {\bibinfo {title} {Phase separation in the
			two-dimensional bosonic hubbard model with ring exchange},\ }\href
	{https://doi.org/10.1103/PhysRevLett.93.110404} {\bibfield  {journal}
		{\bibinfo  {journal} {Phys. Rev. Lett.}\ }\textbf {\bibinfo {volume} {93}},\
		\bibinfo {pages} {110404} (\bibinfo {year} {2004})}\BibitemShut {NoStop}%
	\bibitem [{\citenamefont {Yuan}\ \emph {et~al.}(2020)\citenamefont {Yuan},
		\citenamefont {Chen},\ and\ \citenamefont {Ye}}]{Yuan:2019geh}%
	\BibitemOpen
	\bibfield  {author} {\bibinfo {author} {\bibfnamefont {J.-K.}\ \bibnamefont
			{Yuan}}, \bibinfo {author} {\bibfnamefont {S.~A.}\ \bibnamefont {Chen}},\
		and\ \bibinfo {author} {\bibfnamefont {P.}~\bibnamefont {Ye}},\ }\bibfield
	{title} {\bibinfo {title} {{Fractonic Superfluids}},\ }\href
	{https://doi.org/10.1103/PhysRevResearch.2.023267} {\bibfield  {journal}
		{\bibinfo  {journal} {Phys. Rev. Res.}\ }\textbf {\bibinfo {volume} {2}},\
		\bibinfo {pages} {023267} (\bibinfo {year} {2020})},\ \Eprint
	{https://arxiv.org/abs/1911.02876} {arXiv:1911.02876 [cond-mat.str-el]}
	\BibitemShut {NoStop}%
	\bibitem [{\citenamefont {Chen}\ \emph {et~al.}(2021)\citenamefont {Chen},
		\citenamefont {Yuan},\ and\ \citenamefont {Ye}}]{Chen2021}%
	\BibitemOpen
	\bibfield  {author} {\bibinfo {author} {\bibfnamefont {S.~A.}\ \bibnamefont
			{Chen}}, \bibinfo {author} {\bibfnamefont {J.-K.}\ \bibnamefont {Yuan}},\
		and\ \bibinfo {author} {\bibfnamefont {P.}~\bibnamefont {Ye}},\ }\bibfield
	{title} {\bibinfo {title} {Fractonic superfluids. ii. condensing
			subdimensional particles},\ }\href
	{https://doi.org/10.1103/PhysRevResearch.3.013226} {\bibfield  {journal}
		{\bibinfo  {journal} {Phys. Rev. Research}\ }\textbf {\bibinfo {volume}
			{3}},\ \bibinfo {pages} {013226} (\bibinfo {year} {2021})}\BibitemShut
	{NoStop}%
	\bibitem [{\citenamefont {Gromov}\ \emph {et~al.}(2020)\citenamefont {Gromov},
		\citenamefont {Lucas},\ and\ \citenamefont {Nandkishore}}]{GromovLucas2020}%
	\BibitemOpen
	\bibfield  {author} {\bibinfo {author} {\bibfnamefont {A.}~\bibnamefont
			{Gromov}}, \bibinfo {author} {\bibfnamefont {A.}~\bibnamefont {Lucas}},\ and\
		\bibinfo {author} {\bibfnamefont {R.~M.}\ \bibnamefont {Nandkishore}},\
	}\bibfield  {title} {\bibinfo {title} {Fracton hydrodynamics},\ }\href
	{https://doi.org/10.1103/PhysRevResearch.2.033124} {\bibfield  {journal}
		{\bibinfo  {journal} {Phys. Rev. Research}\ }\textbf {\bibinfo {volume}
			{2}},\ \bibinfo {pages} {033124} (\bibinfo {year} {2020})}\BibitemShut
	{NoStop}%
	\bibitem [{\citenamefont {Aasen}\ \emph {et~al.}(2020)\citenamefont {Aasen},
		\citenamefont {Bulmash}, \citenamefont {Prem}, \citenamefont {Slagle},\ and\
		\citenamefont {Williamson}}]{Aasen:2020zru}%
	\BibitemOpen
	\bibfield  {author} {\bibinfo {author} {\bibfnamefont {D.}~\bibnamefont
			{Aasen}}, \bibinfo {author} {\bibfnamefont {D.}~\bibnamefont {Bulmash}},
		\bibinfo {author} {\bibfnamefont {A.}~\bibnamefont {Prem}}, \bibinfo {author}
		{\bibfnamefont {K.}~\bibnamefont {Slagle}},\ and\ \bibinfo {author}
		{\bibfnamefont {D.~J.}\ \bibnamefont {Williamson}},\ }\bibfield  {title}
	{\bibinfo {title} {{Topological Defect Networks for Fractons of all Types}},\
	}\href {https://doi.org/10.1103/PhysRevResearch.2.043165} {\bibfield
		{journal} {\bibinfo  {journal} {Phys. Rev. Res.}\ }\textbf {\bibinfo {volume}
			{2}},\ \bibinfo {pages} {043165} (\bibinfo {year} {2020})}\BibitemShut
	{NoStop}%
	\bibitem [{\citenamefont {Sonin}(2016)}]{sonin_dynamics_2016}%
	\BibitemOpen
	\bibfield  {author} {\bibinfo {author} {\bibfnamefont {E.~B.}\ \bibnamefont
			{Sonin}},\ }\href@noop {} {\emph {\bibinfo {title} {Dynamics of Quantised
				Vortices in Superfluids.}}}\ (\bibinfo  {publisher} {Cambridge University
		Press},\ \bibinfo {address} {Cambridge},\ \bibinfo {year} {2016})\BibitemShut
	{NoStop}%
	\bibitem [{\citenamefont {Ezawa}(2013)}]{ezawa_quantum_2013}%
	\BibitemOpen
	\bibfield  {author} {\bibinfo {author} {\bibfnamefont {Z.~F.}\ \bibnamefont
			{Ezawa}},\ }\href@noop {} {\emph {\bibinfo {title} {Quantum Hall effects}}}\
	(\bibinfo  {publisher} {World Scientific},\ \bibinfo {year}
	{2013})\BibitemShut {NoStop}%
	\bibitem [{\citenamefont {Turner}\ \emph {et~al.}(2010)\citenamefont {Turner},
		\citenamefont {Vitelli},\ and\ \citenamefont {Nelson}}]{Turner2010}%
	\BibitemOpen
	\bibfield  {author} {\bibinfo {author} {\bibfnamefont {A.~M.}\ \bibnamefont
			{Turner}}, \bibinfo {author} {\bibfnamefont {V.}~\bibnamefont {Vitelli}},\
		and\ \bibinfo {author} {\bibfnamefont {D.~R.}\ \bibnamefont {Nelson}},\
	}\bibfield  {title} {\bibinfo {title} {Vortices on curved surfaces},\ }\href
	{https://doi.org/10.1103/RevModPhys.82.1301} {\bibfield  {journal} {\bibinfo
			{journal} {Rev. Mod. Phys.}\ }\textbf {\bibinfo {volume} {82}},\ \bibinfo
		{pages} {1301} (\bibinfo {year} {2010})}\BibitemShut {NoStop}%
	\bibitem [{\citenamefont {Bertoldi}\ \emph {et~al.}(2017)\citenamefont
		{Bertoldi}, \citenamefont {Vitelli}, \citenamefont {Christensen},\ and\
		\citenamefont {van Hecke}}]{Bertoldi2017}%
	\BibitemOpen
	\bibfield  {author} {\bibinfo {author} {\bibfnamefont {K.}~\bibnamefont
			{Bertoldi}}, \bibinfo {author} {\bibfnamefont {V.}~\bibnamefont {Vitelli}},
		\bibinfo {author} {\bibfnamefont {J.}~\bibnamefont {Christensen}},\ and\
		\bibinfo {author} {\bibfnamefont {M.}~\bibnamefont {van Hecke}},\ }\bibfield
	{title} {\bibinfo {title} {Flexible mechanical metamaterials},\ }\bibfield
	{journal} {\bibinfo  {journal} {Nature Reviews Materials}\ }\textbf {\bibinfo
		{volume} {2}},\ \href {https://doi.org/10.1038/natrevmats.2017.66}
	{10.1038/natrevmats.2017.66} (\bibinfo {year} {2017})\BibitemShut {NoStop}%
	\bibitem [{\citenamefont {Kleinert}(1983)}]{kleinert1983double}%
	\BibitemOpen
	\bibfield  {author} {\bibinfo {author} {\bibfnamefont {H.}~\bibnamefont
			{Kleinert}},\ }\bibfield  {title} {\bibinfo {title} {Double gauge theory of
			stresses and defects},\ }\href {https://doi.org/10.1016/0375-9601(83)90099-3}
	{\bibfield  {journal} {\bibinfo  {journal} {Physics Letters A}\ }\textbf
		{\bibinfo {volume} {97}},\ \bibinfo {pages} {51} (\bibinfo {year}
		{1983})}\BibitemShut {NoStop}%
	\bibitem [{\citenamefont {{Zaanen}}\ \emph {et~al.}(2004)\citenamefont
		{{Zaanen}}, \citenamefont {{Nussinov}},\ and\ \citenamefont
		{{Mukhin}}}]{Zaanen2004}%
	\BibitemOpen
	\bibfield  {author} {\bibinfo {author} {\bibfnamefont {J.}~\bibnamefont
			{{Zaanen}}}, \bibinfo {author} {\bibfnamefont {Z.}~\bibnamefont
			{{Nussinov}}},\ and\ \bibinfo {author} {\bibfnamefont {S.~I.}\ \bibnamefont
			{{Mukhin}}},\ }\bibfield  {title} {\bibinfo {title} {{Duality in 2 + 1D
				quantum elasticity: superconductivity and quantum nematic order}},\ }\href
	{https://doi.org/10.1016/j.aop.2003.10.003} {\bibfield  {journal} {\bibinfo
			{journal} {Annals of Physics}\ }\textbf {\bibinfo {volume} {310}},\ \bibinfo
		{pages} {181} (\bibinfo {year} {2004})}\BibitemShut {NoStop}%
	\bibitem [{\citenamefont {Cvetkovic}\ \emph {et~al.}(2006)\citenamefont
		{Cvetkovic}, \citenamefont {Nussinov},\ and\ \citenamefont
		{Zaanen}}]{Cvetkovic_2006}%
	\BibitemOpen
	\bibfield  {author} {\bibinfo {author} {\bibfnamefont {V.}~\bibnamefont
			{Cvetkovic}}, \bibinfo {author} {\bibfnamefont {Z.}~\bibnamefont
			{Nussinov}},\ and\ \bibinfo {author} {\bibfnamefont {J.}~\bibnamefont
			{Zaanen}},\ }\bibfield  {title} {\bibinfo {title} {Topological kinematic
			constraints: dislocations and the glide principle},\ }\href
	{https://doi.org/10.1080/14786430600636328} {\bibfield  {journal} {\bibinfo
			{journal} {Philosophical Magazine}\ }\textbf {\bibinfo {volume} {86}},\
		\bibinfo {pages} {2995} (\bibinfo {year} {2006})}\BibitemShut {NoStop}%
	\bibitem [{\citenamefont {Beekman}\ \emph {et~al.}(2017)\citenamefont
		{Beekman}, \citenamefont {Nissinen}, \citenamefont {Wu}, \citenamefont {Liu},
		\citenamefont {Slager}, \citenamefont {Nussinov}, \citenamefont {Cvetkovic},\
		and\ \citenamefont {Zaanen}}]{Beekman2017}%
	\BibitemOpen
	\bibfield  {author} {\bibinfo {author} {\bibfnamefont {A.~J.}\ \bibnamefont
			{Beekman}}, \bibinfo {author} {\bibfnamefont {J.}~\bibnamefont {Nissinen}},
		\bibinfo {author} {\bibfnamefont {K.}~\bibnamefont {Wu}}, \bibinfo {author}
		{\bibfnamefont {K.}~\bibnamefont {Liu}}, \bibinfo {author} {\bibfnamefont
			{R.-J.}\ \bibnamefont {Slager}}, \bibinfo {author} {\bibfnamefont
			{Z.}~\bibnamefont {Nussinov}}, \bibinfo {author} {\bibfnamefont
			{V.}~\bibnamefont {Cvetkovic}},\ and\ \bibinfo {author} {\bibfnamefont
			{J.}~\bibnamefont {Zaanen}},\ }\bibfield  {title} {\bibinfo {title} {Dual
			gauge field theory of quantum liquid crystals in two dimensions},\ }\href
	{https://doi.org/10.1016/j.physrep.2017.03.004} {\bibfield  {journal}
		{\bibinfo  {journal} {Physics Reports}\ }\textbf {\bibinfo {volume} {683}},\
		\bibinfo {pages} {1} (\bibinfo {year} {2017})}\BibitemShut {NoStop}%
	\bibitem [{\citenamefont {Wiegmann}\ and\ \citenamefont
		{Abanov}(2014)}]{Wiegmann:2013hca}%
	\BibitemOpen
	\bibfield  {author} {\bibinfo {author} {\bibfnamefont {P.}~\bibnamefont
			{Wiegmann}}\ and\ \bibinfo {author} {\bibfnamefont {A.~G.}\ \bibnamefont
			{Abanov}},\ }\bibfield  {title} {\bibinfo {title} {{Anomalous Hydrodynamics
				of Two-Dimensional Vortex Fluids}},\ }\href
	{https://doi.org/10.1103/PhysRevLett.113.034501} {\bibfield  {journal}
		{\bibinfo  {journal} {Phys. Rev. Lett.}\ }\textbf {\bibinfo {volume} {113}},\
		\bibinfo {pages} {034501} (\bibinfo {year} {2014})}\BibitemShut {NoStop}%
	\bibitem [{\citenamefont {Watanabe}\ and\ \citenamefont
		{Murayama}(2013)}]{Watanabe2013}%
	\BibitemOpen
	\bibfield  {author} {\bibinfo {author} {\bibfnamefont {H.}~\bibnamefont
			{Watanabe}}\ and\ \bibinfo {author} {\bibfnamefont {H.}~\bibnamefont
			{Murayama}},\ }\bibfield  {title} {\bibinfo {title} {Redundancies in
			nambu-goldstone bosons},\ }\href
	{https://doi.org/10.1103/PhysRevLett.110.181601} {\bibfield  {journal}
		{\bibinfo  {journal} {Phys. Rev. Lett.}\ }\textbf {\bibinfo {volume} {110}},\
		\bibinfo {pages} {181601} (\bibinfo {year} {2013})}\BibitemShut {NoStop}%
	\bibitem [{\citenamefont {Lucas}\ and\ \citenamefont
		{Sur\'owka}(2014)}]{LucasSurowkaVortex}%
	\BibitemOpen
	\bibfield  {author} {\bibinfo {author} {\bibfnamefont {A.}~\bibnamefont
			{Lucas}}\ and\ \bibinfo {author} {\bibfnamefont {P.}~\bibnamefont
			{Sur\'owka}},\ }\bibfield  {title} {\bibinfo {title} {Sound-induced vortex
			interactions in a zero-temperature two-dimensional superfluid},\ }\href
	{https://doi.org/10.1103/PhysRevA.90.053617} {\bibfield  {journal} {\bibinfo
			{journal} {Phys. Rev. A}\ }\textbf {\bibinfo {volume} {90}},\ \bibinfo
		{pages} {053617} (\bibinfo {year} {2014})}\BibitemShut {NoStop}%
	\bibitem [{\citenamefont {Yu}\ and\ \citenamefont
		{Bradley}(2017)}]{YuBradley2017}%
	\BibitemOpen
	\bibfield  {author} {\bibinfo {author} {\bibfnamefont {X.}~\bibnamefont
			{Yu}}\ and\ \bibinfo {author} {\bibfnamefont {A.~S.}\ \bibnamefont
			{Bradley}},\ }\bibfield  {title} {\bibinfo {title} {Emergent non-eulerian
			hydrodynamics of quantum vortices in two dimensions},\ }\href
	{https://doi.org/10.1103/PhysRevLett.119.185301} {\bibfield  {journal}
		{\bibinfo  {journal} {Phys. Rev. Lett.}\ }\textbf {\bibinfo {volume} {119}},\
		\bibinfo {pages} {185301} (\bibinfo {year} {2017})}\BibitemShut {NoStop}%
	\bibitem [{\citenamefont {Moroz}\ \emph {et~al.}(2018)\citenamefont {Moroz},
		\citenamefont {Hoyos}, \citenamefont {Benzoni},\ and\ \citenamefont
		{Son}}]{Moroz:2018noc}%
	\BibitemOpen
	\bibfield  {author} {\bibinfo {author} {\bibfnamefont {S.}~\bibnamefont
			{Moroz}}, \bibinfo {author} {\bibfnamefont {C.}~\bibnamefont {Hoyos}},
		\bibinfo {author} {\bibfnamefont {C.}~\bibnamefont {Benzoni}},\ and\ \bibinfo
		{author} {\bibfnamefont {D.~T.}\ \bibnamefont {Son}},\ }\bibfield  {title}
	{\bibinfo {title} {{Effective field theory of a vortex lattice in a bosonic
				superfluid}},\ }\href {https://doi.org/10.21468/SciPostPhys.5.4.039}
	{\bibfield  {journal} {\bibinfo  {journal} {SciPost Phys.}\ }\textbf
		{\bibinfo {volume} {5}},\ \bibinfo {pages} {039} (\bibinfo {year}
		{2018})}\BibitemShut {NoStop}%
	\bibitem [{\citenamefont {Bogatskiy}(2019)}]{Bogatskiy2019}%
	\BibitemOpen
	\bibfield  {author} {\bibinfo {author} {\bibfnamefont {A.}~\bibnamefont
			{Bogatskiy}},\ }\bibfield  {title} {\bibinfo {title} {Vortex flows on closed
			surfaces},\ }\href {https://doi.org/10.1088/1751-8121/ab4e6a} {\bibfield
		{journal} {\bibinfo  {journal} {Journal of Physics A: Mathematical and
				Theoretical}\ }\textbf {\bibinfo {volume} {52}},\ \bibinfo {pages} {475501}
		(\bibinfo {year} {2019})}\BibitemShut {NoStop}%
	\bibitem [{\citenamefont {Wiegmann}(2019)}]{Wiegmann2019}%
	\BibitemOpen
	\bibfield  {author} {\bibinfo {author} {\bibfnamefont {P.}~\bibnamefont
			{Wiegmann}},\ }\bibfield  {title} {\bibinfo {title} {Quantum hydrodynamics,
			rotating superfluid and gravitational anomaly},\ }\href
	{https://doi.org/10.1134/s1063776119100121} {\bibfield  {journal} {\bibinfo
			{journal} {Journal of Experimental and Theoretical Physics}\ }\textbf
		{\bibinfo {volume} {129}},\ \bibinfo {pages} {642} (\bibinfo {year}
		{2019})}\BibitemShut {NoStop}%
	\bibitem [{\citenamefont {Doshi}\ and\ \citenamefont
		{Gromov}(2021)}]{Doshi2021}%
	\BibitemOpen
	\bibfield  {author} {\bibinfo {author} {\bibfnamefont {D.}~\bibnamefont
			{Doshi}}\ and\ \bibinfo {author} {\bibfnamefont {A.}~\bibnamefont {Gromov}},\
	}\bibfield  {title} {\bibinfo {title} {Vortices as fractons},\ }\bibfield
	{journal} {\bibinfo  {journal} {Communications Physics}\ }\textbf {\bibinfo
		{volume} {4}},\ \href {https://doi.org/10.1038/s42005-021-00540-4}
	{10.1038/s42005-021-00540-4} (\bibinfo {year} {2021})\BibitemShut {NoStop}%
	\bibitem [{\citenamefont {Landau}(1941)}]{Landau1941}%
	\BibitemOpen
	\bibfield  {author} {\bibinfo {author} {\bibfnamefont {L.}~\bibnamefont
			{Landau}},\ }\bibfield  {title} {\bibinfo {title} {Theory of the
			superfluidity of helium ii},\ }\href {https://doi.org/10.1103/PhysRev.60.356}
	{\bibfield  {journal} {\bibinfo  {journal} {Phys. Rev.}\ }\textbf {\bibinfo
			{volume} {60}},\ \bibinfo {pages} {356} (\bibinfo {year} {1941})}\BibitemShut
	{NoStop}%
	\bibitem [{\citenamefont {Dzyaloshinskii}\ and\ \citenamefont
		{Volovick}(1980)}]{Dzyaloshinskii1980}%
	\BibitemOpen
	\bibfield  {author} {\bibinfo {author} {\bibfnamefont {I.}~\bibnamefont
			{Dzyaloshinskii}}\ and\ \bibinfo {author} {\bibfnamefont {G.}~\bibnamefont
			{Volovick}},\ }\bibfield  {title} {\bibinfo {title} {Poisson brackets in
			condensed matter physics},\ }\href
	{https://doi.org/10.1016/0003-4916(80)90119-0} {\bibfield  {journal}
		{\bibinfo  {journal} {Annals of Physics}\ }\textbf {\bibinfo {volume}
			{125}},\ \bibinfo {pages} {67} (\bibinfo {year} {1980})}\BibitemShut
	{NoStop}%
	\bibitem [{\citenamefont {Morrison}\ and\ \citenamefont
		{Greene}(1980)}]{Morrison1980}%
	\BibitemOpen
	\bibfield  {author} {\bibinfo {author} {\bibfnamefont {P.~J.}\ \bibnamefont
			{Morrison}}\ and\ \bibinfo {author} {\bibfnamefont {J.~M.}\ \bibnamefont
			{Greene}},\ }\bibfield  {title} {\bibinfo {title} {Noncanonical hamiltonian
			density formulation of hydrodynamics and ideal magnetohydrodynamics},\ }\href
	{https://doi.org/10.1103/PhysRevLett.45.790} {\bibfield  {journal} {\bibinfo
			{journal} {Phys. Rev. Lett.}\ }\textbf {\bibinfo {volume} {45}},\ \bibinfo
		{pages} {790} (\bibinfo {year} {1980})}\BibitemShut {NoStop}%
	\bibitem [{\citenamefont {Zakharov}\ and\ \citenamefont
		{Kuznetsov}(1997)}]{zakharov_hamiltonian_1997}%
	\BibitemOpen
	\bibfield  {author} {\bibinfo {author} {\bibfnamefont {V.~E.}\ \bibnamefont
			{Zakharov}}\ and\ \bibinfo {author} {\bibfnamefont {E.~A.}\ \bibnamefont
			{Kuznetsov}},\ }\bibfield  {title} {\bibinfo {title} {Hamiltonian formalism
			for nonlinear waves},\ }\href
	{https://doi.org/10.1070/PU1997v040n11ABEH000304} {\bibfield  {journal}
		{\bibinfo  {journal} {Physics-Uspekhi}\ }\textbf {\bibinfo {volume} {40}},\
		\bibinfo {pages} {1087} (\bibinfo {year} {1997})}\BibitemShut {NoStop}%
	\bibitem [{\citenamefont {Son}(2019)}]{Son:2019qlm}%
	\BibitemOpen
	\bibfield  {author} {\bibinfo {author} {\bibfnamefont {D.~T.}\ \bibnamefont
			{Son}},\ }\bibfield  {title} {\bibinfo {title} {{Chiral Metric Hydrodynamics,
				Kelvin Circulation Theorem, and the Fractional Quantum Hall Effect}},\
	}\href@noop {} {\  (\bibinfo {year} {2019})},\ \Eprint
	{https://arxiv.org/abs/1907.07187} {arXiv:1907.07187 [cond-mat.str-el]}
	\BibitemShut {NoStop}%
	\bibitem [{\citenamefont {Gromov}(2019{\natexlab{b}})}]{Gromov_multipole_2019}%
	\BibitemOpen
	\bibfield  {author} {\bibinfo {author} {\bibfnamefont {A.}~\bibnamefont
			{Gromov}},\ }\bibfield  {title} {\bibinfo {title} {Towards classification of
			fracton phases: The multipole algebra},\ }\href
	{https://doi.org/10.1103/PhysRevX.9.031035} {\bibfield  {journal} {\bibinfo
			{journal} {Phys. Rev. X}\ }\textbf {\bibinfo {volume} {9}},\ \bibinfo {pages}
		{031035} (\bibinfo {year} {2019}{\natexlab{b}})}\BibitemShut {NoStop}%
	\bibitem [{\citenamefont {Seiberg}\ and\ \citenamefont
		{Shao}(2021)}]{Seiberg:2020bhn}%
	\BibitemOpen
	\bibfield  {author} {\bibinfo {author} {\bibfnamefont {N.}~\bibnamefont
			{Seiberg}}\ and\ \bibinfo {author} {\bibfnamefont {S.-H.}\ \bibnamefont
			{Shao}},\ }\bibfield  {title} {\bibinfo {title} {{Exotic Symmetries, Duality,
				and Fractons in 2+1-Dimensional Quantum Field Theory}},\ }\href
	{https://doi.org/10.21468/SciPostPhys.10.2.027} {\bibfield  {journal}
		{\bibinfo  {journal} {SciPost Phys.}\ }\textbf {\bibinfo {volume} {10}},\
		\bibinfo {pages} {027} (\bibinfo {year} {2021})}\BibitemShut {NoStop}%
	\bibitem [{\citenamefont {Pretko}(2018)}]{Pretko:2018jbi}%
	\BibitemOpen
	\bibfield  {author} {\bibinfo {author} {\bibfnamefont {M.}~\bibnamefont
			{Pretko}},\ }\bibfield  {title} {\bibinfo {title} {{The Fracton Gauge
				Principle}},\ }\href {https://doi.org/10.1103/PhysRevB.98.115134} {\bibfield
		{journal} {\bibinfo  {journal} {Phys. Rev. B}\ }\textbf {\bibinfo {volume}
			{98}},\ \bibinfo {pages} {115134} (\bibinfo {year} {2018})},\ \Eprint
	{https://arxiv.org/abs/1807.11479} {arXiv:1807.11479 [cond-mat.str-el]}
	\BibitemShut {NoStop}%
	\bibitem [{\citenamefont {Seiberg}(2020)}]{Seiberg:2019vrp}%
	\BibitemOpen
	\bibfield  {author} {\bibinfo {author} {\bibfnamefont {N.}~\bibnamefont
			{Seiberg}},\ }\bibfield  {title} {\bibinfo {title} {{Field Theories With a
				Vector Global Symmetry}},\ }\href
	{https://doi.org/10.21468/SciPostPhys.8.4.050} {\bibfield  {journal}
		{\bibinfo  {journal} {SciPost Phys.}\ }\textbf {\bibinfo {volume} {8}},\
		\bibinfo {pages} {050} (\bibinfo {year} {2020})},\ \Eprint
	{https://arxiv.org/abs/1909.10544} {arXiv:1909.10544 [cond-mat.str-el]}
	\BibitemShut {NoStop}%
	\bibitem [{\citenamefont {Stark}\ and\ \citenamefont
		{Lubensky}(2005)}]{Stark2005}%
	\BibitemOpen
	\bibfield  {author} {\bibinfo {author} {\bibfnamefont {H.}~\bibnamefont
			{Stark}}\ and\ \bibinfo {author} {\bibfnamefont {T.~C.}\ \bibnamefont
			{Lubensky}},\ }\bibfield  {title} {\bibinfo {title} {Poisson bracket approach
			to the dynamics of nematic liquid crystals: The role of spin angular
			momentum},\ }\href {https://doi.org/10.1103/PhysRevE.72.051714} {\bibfield
		{journal} {\bibinfo  {journal} {Phys. Rev. E}\ }\textbf {\bibinfo {volume}
			{72}},\ \bibinfo {pages} {051714} (\bibinfo {year} {2005})}\BibitemShut
	{NoStop}%
	\bibitem [{\citenamefont {Nicolis}\ \emph {et~al.}(2009)\citenamefont
		{Nicolis}, \citenamefont {Rattazzi},\ and\ \citenamefont
		{Trincherini}}]{Nicolis:2008in}%
	\BibitemOpen
	\bibfield  {author} {\bibinfo {author} {\bibfnamefont {A.}~\bibnamefont
			{Nicolis}}, \bibinfo {author} {\bibfnamefont {R.}~\bibnamefont {Rattazzi}},\
		and\ \bibinfo {author} {\bibfnamefont {E.}~\bibnamefont {Trincherini}},\
	}\bibfield  {title} {\bibinfo {title} {{The Galileon as a local modification
				of gravity}},\ }\href {https://doi.org/10.1103/PhysRevD.79.064036} {\bibfield
		{journal} {\bibinfo  {journal} {Phys. Rev. D}\ }\textbf {\bibinfo {volume}
			{79}},\ \bibinfo {pages} {064036} (\bibinfo {year} {2009})}\BibitemShut
	{NoStop}%
	\bibitem [{\citenamefont {Griffin}\ \emph {et~al.}(2015)\citenamefont
		{Griffin}, \citenamefont {Grosvenor}, \citenamefont
		{Ho\ifmmode~\check{r}\else \v{r}\fi{}ava},\ and\ \citenamefont
		{Yan}}]{Griffin:2014bta}%
	\BibitemOpen
	\bibfield  {author} {\bibinfo {author} {\bibfnamefont {T.}~\bibnamefont
			{Griffin}}, \bibinfo {author} {\bibfnamefont {K.~T.}\ \bibnamefont
			{Grosvenor}}, \bibinfo {author} {\bibfnamefont {P.}~\bibnamefont
			{Ho\ifmmode~\check{r}\else \v{r}\fi{}ava}},\ and\ \bibinfo {author}
		{\bibfnamefont {Z.}~\bibnamefont {Yan}},\ }\bibfield  {title} {\bibinfo
		{title} {{Scalar Field Theories with Polynomial Shift Symmetries}},\ }\href
	{https://doi.org/10.1007/s00220-015-2461-2} {\bibfield  {journal} {\bibinfo
			{journal} {Commun. Math. Phys.}\ }\textbf {\bibinfo {volume} {340}},\
		\bibinfo {pages} {985} (\bibinfo {year} {2015})}\BibitemShut {NoStop}%
	\bibitem [{\citenamefont {Pretko}(2017{\natexlab{c}})}]{Pretko:2016lgv}%
	\BibitemOpen
	\bibfield  {author} {\bibinfo {author} {\bibfnamefont {M.}~\bibnamefont
			{Pretko}},\ }\bibfield  {title} {\bibinfo {title} {{Generalized
				Electromagnetism of Subdimensional Particles: A Spin Liquid Story}},\ }\href
	{https://doi.org/10.1103/PhysRevB.96.035119} {\bibfield  {journal} {\bibinfo
			{journal} {Phys. Rev. B}\ }\textbf {\bibinfo {volume} {96}},\ \bibinfo
		{pages} {035119} (\bibinfo {year} {2017}{\natexlab{c}})}\BibitemShut
	{NoStop}%
	\bibitem [{\citenamefont {Halperin}\ and\ \citenamefont
		{Nelson}(1978)}]{PhysRevLett.41.121}%
	\BibitemOpen
	\bibfield  {author} {\bibinfo {author} {\bibfnamefont {B.~I.}\ \bibnamefont
			{Halperin}}\ and\ \bibinfo {author} {\bibfnamefont {D.~R.}\ \bibnamefont
			{Nelson}},\ }\bibfield  {title} {\bibinfo {title} {Theory of two-dimensional
			melting},\ }\href {https://doi.org/10.1103/PhysRevLett.41.121} {\bibfield
		{journal} {\bibinfo  {journal} {Phys. Rev. Lett.}\ }\textbf {\bibinfo
			{volume} {41}},\ \bibinfo {pages} {121} (\bibinfo {year} {1978})}\BibitemShut
	{NoStop}%
	\bibitem [{\citenamefont {Karch}\ and\ \citenamefont
		{Raz}(2021)}]{Karch:2020yuy}%
	\BibitemOpen
	\bibfield  {author} {\bibinfo {author} {\bibfnamefont {A.}~\bibnamefont
			{Karch}}\ and\ \bibinfo {author} {\bibfnamefont {A.}~\bibnamefont {Raz}},\
	}\bibfield  {title} {\bibinfo {title} {{Reduced Conformal Symmetry}},\ }\href
	{https://doi.org/10.1007/JHEP04(2021)182} {\bibfield  {journal} {\bibinfo
			{journal} {JHEP}\ }\textbf {\bibinfo {volume} {04}},\ \bibinfo {pages}
		{182}},\ \Eprint {https://arxiv.org/abs/2009.12308} {arXiv:2009.12308
		[hep-th]} \BibitemShut {NoStop}%
	\bibitem [{\citenamefont {Kleman}\ and\ \citenamefont
		{Friedel}(2008)}]{Kleman2008review}%
	\BibitemOpen
	\bibfield  {author} {\bibinfo {author} {\bibfnamefont {M.}~\bibnamefont
			{Kleman}}\ and\ \bibinfo {author} {\bibfnamefont {J.}~\bibnamefont
			{Friedel}},\ }\bibfield  {title} {\bibinfo {title} {Disclinations,
			dislocations, and continuous defects: A reappraisal},\ }\href
	{https://doi.org/10.1103/RevModPhys.80.61} {\bibfield  {journal} {\bibinfo
			{journal} {Rev. Mod. Phys.}\ }\textbf {\bibinfo {volume} {80}},\ \bibinfo
		{pages} {61} (\bibinfo {year} {2008})}\BibitemShut {NoStop}%
	\bibitem [{\citenamefont {{Shankar}}\ \emph {et~al.}(2020)\citenamefont
		{{Shankar}}, \citenamefont {{Souslov}}, \citenamefont {{Bowick}},
		\citenamefont {{Marchetti}},\ and\ \citenamefont {{Vitelli}}}]{Shankar2020}%
	\BibitemOpen
	\bibfield  {author} {\bibinfo {author} {\bibfnamefont {S.}~\bibnamefont
			{{Shankar}}}, \bibinfo {author} {\bibfnamefont {A.}~\bibnamefont
			{{Souslov}}}, \bibinfo {author} {\bibfnamefont {M.~J.}\ \bibnamefont
			{{Bowick}}}, \bibinfo {author} {\bibfnamefont {M.~C.}\ \bibnamefont
			{{Marchetti}}},\ and\ \bibinfo {author} {\bibfnamefont {V.}~\bibnamefont
			{{Vitelli}}},\ }\bibfield  {title} {\bibinfo {title} {{Topological active
				matter}},\ }\href@noop {} {\  (\bibinfo {year} {2020})},\ \Eprint
	{https://arxiv.org/abs/2010.00364} {arXiv:2010.00364 [cond-mat.soft]}
	\BibitemShut {NoStop}%
	\bibitem [{\citenamefont {Mura}(1963)}]{Mura1963}%
	\BibitemOpen
	\bibfield  {author} {\bibinfo {author} {\bibfnamefont {T.}~\bibnamefont
			{Mura}},\ }\bibfield  {title} {\bibinfo {title} {Continuous distribution of
			moving dislocations},\ }\href {https://doi.org/10.1080/14786436308213841}
	{\bibfield  {journal} {\bibinfo  {journal} {Philosophical Magazine}\ }\textbf
		{\bibinfo {volume} {8}},\ \bibinfo {pages} {843} (\bibinfo {year}
		{1963})}\BibitemShut {NoStop}%
	\bibitem [{\citenamefont {Kosevich}(1965)}]{Kosevich_1965}%
	\BibitemOpen
	\bibfield  {author} {\bibinfo {author} {\bibfnamefont {A.~M.}\ \bibnamefont
			{Kosevich}},\ }\bibfield  {title} {\bibinfo {title} {{Dynamical} {Theory}
			{of} {Dislocations}},\ }\href
	{https://doi.org/10.1070/pu1965v007n06abeh003688} {\bibfield  {journal}
		{\bibinfo  {journal} {Soviet Physics Uspekhi}\ }\textbf {\bibinfo {volume}
			{7}},\ \bibinfo {pages} {837} (\bibinfo {year} {1965})}\BibitemShut {NoStop}%
	\bibitem [{\citenamefont {Julia}\ and\ \citenamefont
		{Toulouse}(1979)}]{Julia1979}%
	\BibitemOpen
	\bibfield  {author} {\bibinfo {author} {\bibfnamefont {B.}~\bibnamefont
			{Julia}}\ and\ \bibinfo {author} {\bibfnamefont {G.}~\bibnamefont
			{Toulouse}},\ }\bibfield  {title} {\bibinfo {title} {The many-defect problem
			: gauge-like variables for ordered media containing defects},\ }\href
	{https://doi.org/10.1051/jphyslet:019790040016039500} {\bibfield  {journal}
		{\bibinfo  {journal} {Journal de Physique Lettres}\ }\textbf {\bibinfo
			{volume} {40}},\ \bibinfo {pages} {395} (\bibinfo {year} {1979})}\BibitemShut
	{NoStop}%
	\bibitem [{\citenamefont {Acharya}(2001)}]{Acharya2001}%
	\BibitemOpen
	\bibfield  {author} {\bibinfo {author} {\bibfnamefont {A.}~\bibnamefont
			{Acharya}},\ }\bibfield  {title} {\bibinfo {title} {A model of crystal
			plasticity based on the theory of continuously distributed dislocations},\
	}\href {https://doi.org/10.1016/s0022-5096(00)00060-0} {\bibfield  {journal}
		{\bibinfo  {journal} {Journal of the Mechanics and Physics of Solids}\
		}\textbf {\bibinfo {volume} {49}},\ \bibinfo {pages} {761} (\bibinfo {year}
		{2001})}\BibitemShut {NoStop}%
	\bibitem [{\citenamefont {Fressengeas}\ \emph {et~al.}(2011)\citenamefont
		{Fressengeas}, \citenamefont {Taupin},\ and\ \citenamefont
		{Capolungo}}]{FRESSENGEAS20113499}%
	\BibitemOpen
	\bibfield  {author} {\bibinfo {author} {\bibfnamefont {C.}~\bibnamefont
			{Fressengeas}}, \bibinfo {author} {\bibfnamefont {V.}~\bibnamefont
			{Taupin}},\ and\ \bibinfo {author} {\bibfnamefont {L.}~\bibnamefont
			{Capolungo}},\ }\bibfield  {title} {\bibinfo {title} {An elasto-plastic
			theory of dislocation and disclination fields},\ }\href
	{https://doi.org/https://doi.org/10.1016/j.ijsolstr.2011.09.002} {\bibfield
		{journal} {\bibinfo  {journal} {International Journal of Solids and
				Structures}\ }\textbf {\bibinfo {volume} {48}},\ \bibinfo {pages} {3499}
		(\bibinfo {year} {2011})}\BibitemShut {NoStop}%
	\bibitem [{\citenamefont {T{\'o}th}\ \emph {et~al.}(2002)\citenamefont
		{T{\'o}th}, \citenamefont {Denniston},\ and\ \citenamefont
		{Yeomans}}]{Toth2002}%
	\BibitemOpen
	\bibfield  {author} {\bibinfo {author} {\bibfnamefont {G.}~\bibnamefont
			{T{\'o}th}}, \bibinfo {author} {\bibfnamefont {C.}~\bibnamefont
			{Denniston}},\ and\ \bibinfo {author} {\bibfnamefont {J.~M.}\ \bibnamefont
			{Yeomans}},\ }\bibfield  {title} {\bibinfo {title} {Hydrodynamics of
			topological defects in nematic liquid crystals},\ }\href
	{https://doi.org/10.1103/PhysRevLett.88.105504} {\bibfield  {journal}
		{\bibinfo  {journal} {Phys. Rev. Lett.}\ }\textbf {\bibinfo {volume} {88}},\
		\bibinfo {pages} {105504} (\bibinfo {year} {2002})}\BibitemShut {NoStop}%
	\bibitem [{\citenamefont {Tang}\ and\ \citenamefont
		{Selinger}(2019)}]{Tang2019}%
	\BibitemOpen
	\bibfield  {author} {\bibinfo {author} {\bibfnamefont {X.}~\bibnamefont
			{Tang}}\ and\ \bibinfo {author} {\bibfnamefont {J.~V.}\ \bibnamefont
			{Selinger}},\ }\bibfield  {title} {\bibinfo {title} {Theory of defect motion
			in 2d passive and active nematic liquid crystals},\ }\href
	{https://doi.org/10.1039/C8SM01901K} {\bibfield  {journal} {\bibinfo
			{journal} {Soft Matter}\ }\textbf {\bibinfo {volume} {15}},\ \bibinfo {pages}
		{587} (\bibinfo {year} {2019})}\BibitemShut {NoStop}%
	\bibitem [{\citenamefont {{Acharya}}(2021)}]{Acharya2021}%
	\BibitemOpen
	\bibfield  {author} {\bibinfo {author} {\bibfnamefont {A.}~\bibnamefont
			{{Acharya}}},\ }\bibfield  {title} {\bibinfo {title} {{An action for
				nonlinear dislocation dynamics}},\ }\href@noop {} {\  (\bibinfo {year}
		{2021})},\ \Eprint {https://arxiv.org/abs/2104.12568} {arXiv:2104.12568
		[cond-mat.mtrl-sci]} \BibitemShut {NoStop}%
	\bibitem [{\citenamefont {Harris}\ and\ \citenamefont
		{Scriven}(1971)}]{Harris:1971}%
	\BibitemOpen
	\bibfield  {author} {\bibinfo {author} {\bibfnamefont {W.~F.}\ \bibnamefont
			{Harris}}\ and\ \bibinfo {author} {\bibfnamefont {L.~E.}\ \bibnamefont
			{Scriven}},\ }\bibfield  {title} {\bibinfo {title} {Intrinsic disclinations
			as dislocation sources and sinks in surface crystals},\ }\href
	{https://doi.org/10.1063/1.1660731} {\bibfield  {journal} {\bibinfo
			{journal} {Journal of Applied Physics}\ }\textbf {\bibinfo {volume} {42}},\
		\bibinfo {pages} {3309} (\bibinfo {year} {1971})}\BibitemShut {NoStop}%
	\bibitem [{\citenamefont {deWit}(1971)}]{deWit:1971}%
	\BibitemOpen
	\bibfield  {author} {\bibinfo {author} {\bibfnamefont {R.}~\bibnamefont
			{deWit}},\ }\bibfield  {title} {\bibinfo {title} {Relation between
			dislocations and disclinations},\ }\href {https://doi.org/10.1063/1.1660730}
	{\bibfield  {journal} {\bibinfo  {journal} {Journal of Applied Physics}\
		}\textbf {\bibinfo {volume} {42}},\ \bibinfo {pages} {3304} (\bibinfo {year}
		{1971})}\BibitemShut {NoStop}%
	\bibitem [{\citenamefont {Gutkin}\ \emph {et~al.}(2003)\citenamefont {Gutkin},
		\citenamefont {Ovid'ko},\ and\ \citenamefont {Skiba}}]{GUTKIN200373}%
	\BibitemOpen
	\bibfield  {author} {\bibinfo {author} {\bibfnamefont {M.}~\bibnamefont
			{Gutkin}}, \bibinfo {author} {\bibfnamefont {I.}~\bibnamefont {Ovid'ko}},\
		and\ \bibinfo {author} {\bibfnamefont {N.}~\bibnamefont {Skiba}},\ }\bibfield
	{title} {\bibinfo {title} {Transformations of grain boundaries due to
			disclination motion and emission of dislocation pairs},\ }\href
	{https://doi.org/https://doi.org/10.1016/S0921-5093(02)00107-7} {\bibfield
		{journal} {\bibinfo  {journal} {Materials Science and Engineering: A}\
		}\textbf {\bibinfo {volume} {339}},\ \bibinfo {pages} {73} (\bibinfo {year}
		{2003})}\BibitemShut {NoStop}%
	\bibitem [{\citenamefont {Desai}\ and\ \citenamefont
		{Kapral}(2009)}]{desai_kapral_2009}%
	\BibitemOpen
	\bibfield  {author} {\bibinfo {author} {\bibfnamefont {R.~C.}\ \bibnamefont
			{Desai}}\ and\ \bibinfo {author} {\bibfnamefont {R.}~\bibnamefont {Kapral}},\
	}\href {https://doi.org/10.1017/CBO9780511609725} {\emph {\bibinfo {title}
			{Dynamics of Self-Organized and Self-Assembled Structures}}}\ (\bibinfo
	{publisher} {Cambridge University Press},\ \bibinfo {year}
	{2009})\BibitemShut {NoStop}%
\end{thebibliography}
\end{document}